\documentclass[11pt]{article}

\usepackage[final]{acl}

\usepackage{times}
\usepackage{latexsym}
\usepackage[most]{tcolorbox}
\usepackage[T1]{fontenc}
\usepackage{makecell}
\usepackage[utf8]{inputenc}

\usepackage{microtype}

\usepackage{inconsolata}

\usepackage{graphicx}
\usepackage{subcaption}
\usepackage{amsmath}
\usepackage{amssymb}
\usepackage[ruled,vlined,linesnumbered]{algorithm2e}
\usepackage{booktabs}
\usepackage{multirow}
\usepackage[table]{xcolor}
\usepackage{xcolor}
\usepackage{tabularray}
\UseTblrLibrary{booktabs}
\usepackage{multirow}
\usepackage[most]{tcolorbox}
\tcbuselibrary{listings,breakable}
\usepackage[ruled,vlined,linesnumbered]{algorithm2e}

\usepackage[ruled,linesnumbered]{algorithm2e}
\usepackage{enumitem}

\SetAlgoNoLine
\SetKwInput{KwIn}{Require}
\SetKwInput{KwOut}{Return}
\SetKwComment{tcp}{$\triangleright$\ }{}
\SetAlgoNlRelativeSize{-1}
\DontPrintSemicolon

\title{PIVOT: Preference-based Intervention Vectors for Pedagogical Tutor Steering}

\author{
  Fares Fawzi, 
  Jiaxu Zhao,
  Tanya Nazaretsky,
  Tanja Käser\\
  EPFL \\
  \texttt{\{firstname.lastname\}@epfl.ch} \\
}

\begin{document}
\maketitle
\begin{abstract}
LLMs are increasingly used for conversational tutoring, but effective tutoring requires more than correct answers. Tutors must choose when to scaffold reasoning, hint, give feedback, explain, or invite reflection. Existing prompting and training methods improve pedagogical alignment, but lack reliable inference-time control over pedagogical strategies.
We introduce PIVOT, an activation-steering framework that learns preference-based intervention vectors online for frozen LLM tutors. PIVOT uses a seven-category tutor-move taxonomy and a generate–label–optimise loop, where a human-validated LLM judge identifies target and confusable non-target moves to construct preference pairs for multi-layer residual-stream steering. Across held-out and out-of-domain tutoring data, PIVOT controls tutor moves while preserving relevance and fluency, and its directions can be scaled, transferred, and composed at inference time. In a user study with 30 teachers, 73.3\% of participants preferred steered conversations over neutral baseline interactions using the same prompt, and rated the controls as clear, usable, and pedagogically meaningful.

% We introduce PIVOT, an activation-steering framework that learns preference-based intervention vectors online to steer frozen LLM tutors toward explicit pedagogical moves. PIVOT derives a seven-category tutor-move taxonomy and learns steering vectors through an online generate-label-optimise loop. 
% A human-validated LLM judge labels target and confusable non-target moves in the model's own generations, allowing PIVOT to construct preference pairs and optimise multi-layer residual-stream steering. Across held-out and out-of-domain tutoring data, PIVOT achieves target-move control while preserving relevance and fluency, and its learned directions can be scaled, transferred, and composed at inference time. 

\end{abstract}

\section{Introduction}
% What is the problem and why is it important
% Large language models (LLMs) are increasingly being developed for educational applications, including question generation \cite{scaria2024automated,2024_Hang}, automatic feedback generation \cite{fawzi2026refinerealworldexplorationinteractive,10.1007/978-3-031-72315-5_20}, and interactive explanation \cite{fawzi-etal-2025-scribe}. Particularly, conversational tutoring has been explored in prior work \cite{sonkar-etal-2023-class,NEURIPS2024_9bae399d,dinucu-jianu-etal-2025-problem}, as a promising application because it can provide scalable, individualised support to learners \cite{jurenka2025responsibledevelopmentgenerativeai,puech-etal-2025-towards}. Effective tutoring requires more than correct answers: models must identify student errors, choose appropriate pedagogical strategies, and guide learners toward constructing solutions, since productive challenge can support more durable learning than immediate answer provision \cite{maurya-etal-2025-unifying,dinucu-jianu-etal-2025-problem,bjork2011making}. This task is difficult for LLMs because they are optimised for assistant-like helpfulness rather than pedagogical behaviour \cite{puech-etal-2025-towards,sonkar-etal-2024-pedagogical}. As a result, LLM tutors may reveal answers too early \cite{zhao2026evaluatinganswerleakagerobustness}, provide too much support, or fail to sustain productive multi-turn tutoring interactions \cite{puech-etal-2025-towards}.
Large language models (LLMs) are increasingly used for educational tasks such as feedback generation \cite{fawzi2026refinerealworldexplorationinteractive,10.1007/978-3-031-72315-5_20}, explanation \cite{fawzi-etal-2025-scribe}, and conversational tutoring.
Conversational tutoring \cite{sonkar-etal-2023-class,NEURIPS2024_9bae399d,dinucu-jianu-etal-2025-problem} is especially promising because it can provide scalable, individualised support to learners \cite{jurenka2025responsibledevelopmentgenerativeai,puech-etal-2025-towards}. However, effective tutoring requires more than producing correct answers: tutors must guide students through productive challenge rather than immediate solution delivery \cite{maurya-etal-2025-unifying,dinucu-jianu-etal-2025-problem,bjork2011making}. Since LLMs are optimised for assistant-style helpfulness rather than pedagogy \cite{puech-etal-2025-towards,sonkar-etal-2024-pedagogical}, they often reveal answers too early, provide excessive support, or fail to sustain productive multi-turn tutoring \cite{zhao2026evaluatinganswerleakagerobustness,puech-etal-2025-towards}.

\begin{figure}[t]
    \centering
    \includegraphics[width=\columnwidth]{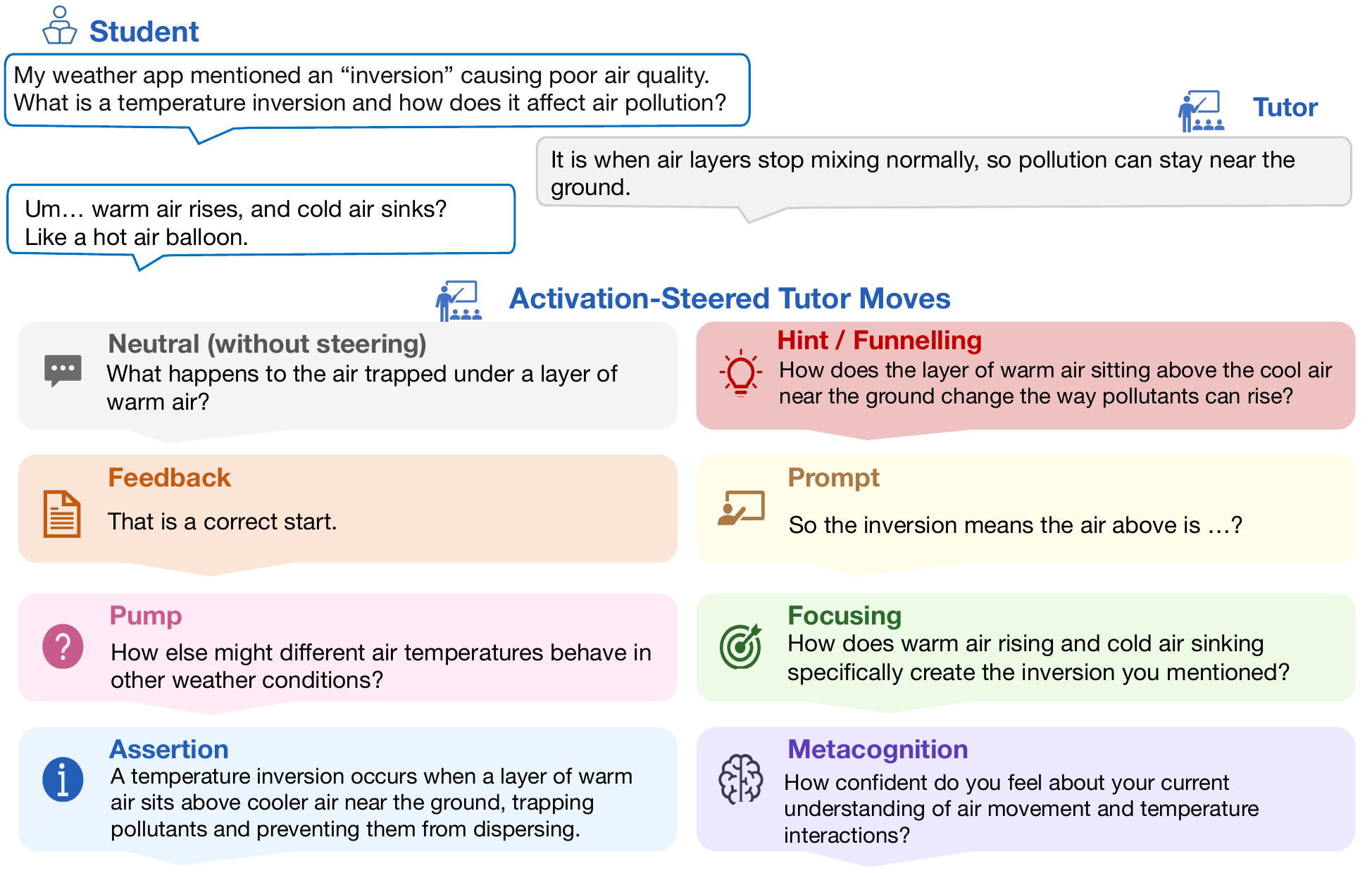}
    \caption{
    PIVOT steers a frozen LLM toward different pedagogical moves for the same student question.
    % while preserving contextual relevance.
    }
    \label{fig:pull}
\end{figure}

% Consequently, a growing body of work has sought to improve the pedagogical alignment in LLM tutors using prompting, supervised fine-tuning, and optimisation-based alignment. Prompting-based methods encode pedagogical policies in context, including step-by-step scaffolding \cite{sonkar-etal-2023-class}, adaptation to student states \cite{10.1145/3613905.3651122}, course-plan updates from interaction history \cite{10.1145/3627673.3679665}, and strategy-level steering through student-state transitions \cite{puech-etal-2025-towards}. Alternatively, training-based methods embed pedagogical behaviour into the model through supervised fine-tuning on tutoring dialogues \cite{jurenka2025responsibledevelopmentgenerativeai,macina-etal-2023-mathdial,sonkar-etal-2023-class}, DPO-style preference optimisation over tutor utterances \cite{10.1007/978-3-031-98414-3_18,rafailov2023direct}, or reinforcement learning from simulated student-tutor interactions \cite{dinucu-jianu-etal-2025-problem}. While these approaches improve alignment, they have different limitations for control. Prompting keeps behaviour externally specified, but can fail across contexts or under adversarial student requests \cite{zhao2026evaluatinganswerleakagerobustness}. Training-based methods can make behaviour more stable, but do not expose tutor moves as controls that can be selected, scaled, or combined at inference time.

Consequently, prior work aligns LLM tutors through either prompting or training-based alignment. Prompting methods specify pedagogical strategies in context, including scaffolding, student-state adaptation, and tutoring policies \cite{sonkar-etal-2023-class,10.1145/3613905.3651122,10.1145/3627673.3679665,puech-etal-2025-towards}. Training-based approaches instead encode pedagogical behaviour through supervised fine-tuning (SFT), preference optimisation, or reinforcement learning (RL) \cite{jurenka2025responsibledevelopmentgenerativeai,macina-etal-2023-mathdial,10.1007/978-3-031-98414-3_18,rafailov2023direct,dinucu-jianu-etal-2025-problem}. However, both limit controllability differently: prompting is adaptable but fragile under distribution shifts or adversarial requests \cite{zhao2026evaluatinganswerleakagerobustness}, while training-based alignment produces more stable behaviour at the cost of explicit inference-time control over tutor moves.

Activation steering enables inference-time control over LLM behaviour by injecting steering vectors into residual-stream activations, without modifying prompts or model weights \cite{rimsky-etal-2024-steering,cao2024personalized,10.5555/3737916.3742333}. Prior work derives such vectors from behavioural contrasts \cite{turner2024steering,rimsky-etal-2024-steering} or preference optimisation \cite{cao2024personalized}, and recent tutoring work applies them to tutor personas and interaction styles \cite{lee2026lettingtutorpersonasspeak}. However, steering pedagogical tutor moves presents two challenges. First, tutor moves are not naturally binary: behaviours such as feedback, hinting, and direct explanation often overlap, leaving no clear contrastive negative class. Second, tutoring research lacks a unified taxonomy or labelled dataset across tutoring systems, classroom discourse, human tutoring, and LLM tutoring research \cite{Graesser2004,alic-etal-2022-computationally,LIN2022194,OCONNOR2019166,macina-etal-2023-mathdial}. Consequently, there is no common supervision framework for learning move-specific steering representations.

To address these challenges, we propose \textbf{PIVOT}, an online preference-based steering framework for controllable tutoring. PIVOT derives a seven-category tutor-move taxonomy and learns move-specific residual-stream steering vectors for a frozen LLM through a generate--judge--optimise loop. A human-validated LLM judge labels target and confusable non-target moves, while relevance and fluency judges filter low-quality generations. These judgements form preference pairs, allowing vectors to learn from model confusions without pre-labelled tutoring data. At inference time, steering vectors can be activated, scaled, and combined to control pedagogical behaviour without updating model weights. PIVOT steers the same tutoring context toward different moves while preserving relevance (see Figure~\ref{fig:pull}). We evaluate PIVOT through offline LLM-based evaluation and a user study with 30 teachers. PIVOT reliably controls tutor moves while preserving relevance and fluency, generalises to unseen tutoring data, and transfers to a non-instruction-tuned base model. In our user study, 73.3\% of teachers preferred steered conversations over a neutral baseline and rated the controls as pedagogically meaningful and interpretable.
Our main contributions are:

\begin{itemize}[leftmargin=*,topsep=1pt, partopsep=0pt, parsep=0pt, itemsep=1pt]
    \item \textbf{We propose PIVOT, an inference-time framework for controllable pedagogical steering} that learns multi-layer residual-stream steering vectors through online preference optimisation over the model’s own tutor-move confusions.

    \item \textbf{We introduce a seven-category taxonomy of pedagogical tutor moves} and develop a human-validated LLM-based evaluation pipeline for tutor-move, relevance, and fluency assessment.

    \item \textbf{We analyse the controllability and robustness of pedagogical steering}, characterising how steering strength, intervention layers, datasets, and model variants affect tutor-move control, relevance, and fluency.

    \item \textbf{We conduct a user study with 30 teachers} showing that controllable tutor moves are pedagogically interpretable and preferred over neutral tutoring behaviour in a teacher-facing interface. \\
\end{itemize} 

\noindent We provide our implementation at \url{https://anonymous.4open.science/r/PIVOT}. 
%For the camera-ready version, we will label the publicly available datasets used in this work and release them with our tutor-move taxonomy.

\section{Related Work}

\begin{figure*}[t]
  \centering
  \includegraphics[width=\linewidth]{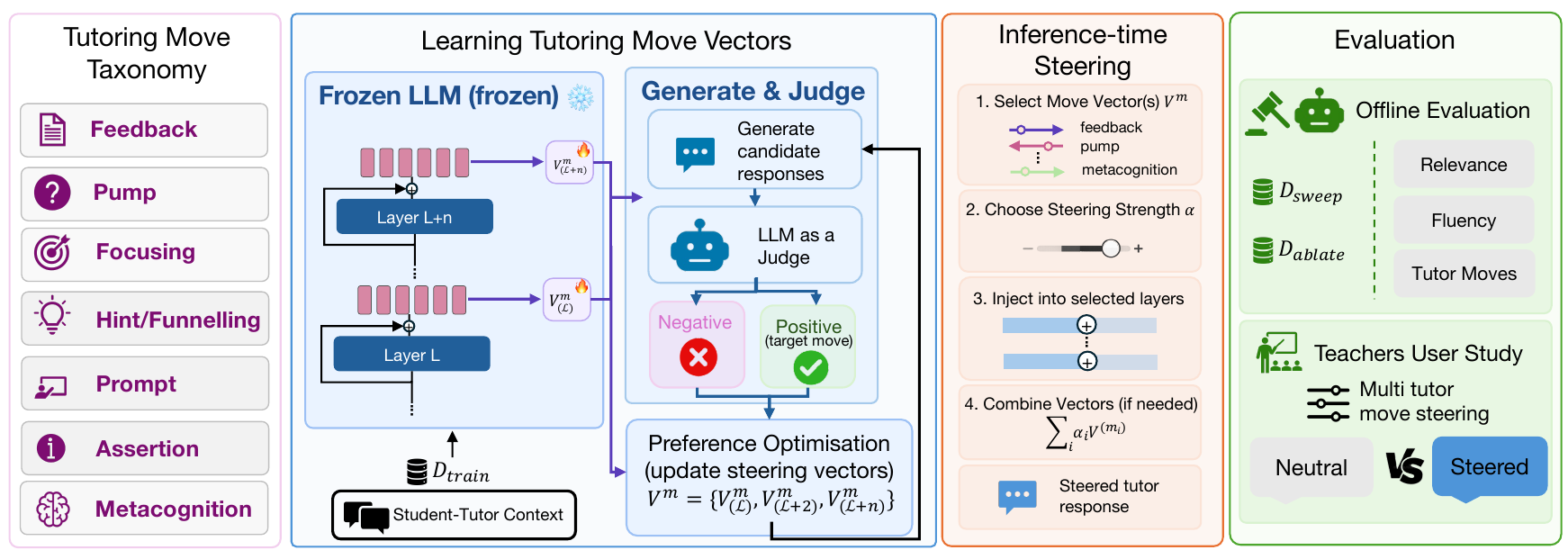}
% \caption{\textbf{PIVOT pipeline and evaluation.} The PIVOT pipeline has three stages: defining a tutoring move taxonomy, learning move-specific residual-stream steering vectors for a frozen LLM from rubric-judged positive and negative generations, and using the learned vectors for inference-time steering. The evaluation component assesses the pipeline with offline LLM-judge metrics and a teacher user study.}
\caption{\textbf{PIVOT pipeline and evaluation.} PIVOT defines a tutor-move taxonomy, learns move-specific residual-stream steering vectors from rubric-judged generations, and applies them for inference-time control. Evaluation uses offline LLM-judge metrics and a user study with teachers.}
  \vspace{-4mm}
  \label{fig:method_fig}
\end{figure*}

\noindent \textbf{LLM Tutors and Pedagogical Alignment.}
Effective tutors guide students through questions, hints, feedback, and adaptive support \cite{LIN2022194,bjork2011making}. In contrast, LLMs are optimised for helpful answer generation rather than pedagogy \cite{2022.EDM-short-papers.54,macina-etal-2023-opportunities}, often leading to premature answer disclosure or incorrect feedback \cite{macina-etal-2023-mathdial,10.1007/978-3-031-98414-3_23}.

Prior work improves pedagogical alignment through prompting or training-based approaches. Prompting methods condition tutoring behaviour on student states, interaction history, or predefined pedagogical structures. For example, ChatTutor updates learning histories through interaction modules and memory \cite{10.1145/3627673.3679665}, while StratL and MWPTutor use predefined tutoring policies \cite{puech-etal-2025-towards,10.1145/3657604.3662041}. However, prompting-based systems remain context-sensitive and vulnerable to adversarial requests \cite{jurenka2025responsibledevelopmentgenerativeai,zhao2026evaluatinganswerleakagerobustness}.

Training-based approaches instead encode pedagogy through SFT, preference optimisation, or RL on tutoring interactions. MathDial fine-tunes on semi-synthetic teacher--student dialogues \cite{macina-etal-2023-mathdial}, Book2Dial and TutorChat use synthetic textbook-grounded dialogues \cite{wang-etal-2024-book2dial,pmlr-v235-chevalier24a}, and LearnLM uses mixed human, synthetic, and teacher-authored pedagogical data \cite{jurenka2025responsibledevelopmentgenerativeai}. Preference- and reward-based methods further optimise tutoring through DPO or RL over simulated interactions \cite{10.1007/978-3-031-98414-3_18,dinucu-jianu-etal-2025-problem}. Although effective, these methods require costly training and provide limited inference-time control.

A further challenge is that tutor-move taxonomies are fragmented across tutoring systems, classroom discourse, and human tutoring research. Prior work studies scripted tutoring moves such as hints, prompts, and feedback \cite{Graesser2004}, classroom questioning \cite{herbel2005questioning,hagenah2018funneling,alic-etal-2022-computationally,OCONNOR2019166}, and dialogue acts in tutoring dialogues \cite{LIN2022194}. However, these taxonomies differ in setting and granularity, providing no unified label space for controllable pedagogical steering.

\vspace{1mm} \noindent \textbf{Activation Steering.}
Activation steering controls LLM behaviour by intervening on hidden activations at inference time, without modifying prompts or model weights \cite{turner2024steering,zou2023representation,li2023inference,rimsky-etal-2024-steering,cao2024personalized}. Early work extracted latent vectors through per-example optimisation \cite{subramani-etal-2022-extracting}. Activation Addition instead computes residual-stream directions from paired prompts differing in a target concept \cite{turner2024steering}, while \citet{li2023inference} steer truthfulness by intervening on activations associated with selected attention heads. Later methods improve robustness through averaged behavioural contrasts (CAA) \cite{rimsky-etal-2024-steering} and preference-optimised steering vectors (BiPO) \cite{cao2024personalized}. However, existing methods typically rely on binary contrasts, such as truthful versus untruthful or harmless versus harmful generations. Pedagogical tutor moves are less clearly separable: tutors may scaffold reasoning, provide partial guidance, invite reflection, or directly explain within the same interaction. Moreover, tutoring research lacks unified tutor-move labels and clean target-opposite contrasts, complicating pedagogical steering.

\section{Methods}
Our goal is to enable inference-time control over LLM tutoring behaviour through steerable pedagogical moves. PIVOT consists of three stages followed by evaluation (Figure~\ref{fig:method_fig}). First, we define a seven-category tutor-move taxonomy. We then use a frozen LLM and a human-validated LLM judge to learn move-specific residual-stream steering vectors from rubric-grounded positive and negative generations using preference optimisation. At inference time, the vectors can be activated, scaled, and combined. Finally, we evaluate PIVOT both offline and in a user study with teachers.

%%%%%%%%%%%%%%%%%%%
\subsection{Tutoring Move Taxonomy}
We derived a seven-category tutoring move taxonomy from prior work on intelligent tutoring, classroom discourse, productive classroom talk, and tutoring-dialogue mining. Table \ref{tab:tutoring-move-taxonomy} summarises each move, its definition, examples, type, and grounding in prior work. The taxonomy distinguishes scaffolding moves, which support reasoning without revealing the answer, from answer-revealing moves, which provide increasing amounts of solution information. Finally, a reflection move encourages students to reflect on their feelings.
\vspace{-0.08cm}

\newcommand{\rotlabel}[1]{%
  \makebox[0pt][c]{%
    \raisebox{0pt}[0pt][0pt]{%
      \tiny
      \rotatebox[origin=c]{90}{\textbf{#1}}%
    }%
  }%
}

\begin{table*}[t]
\centering
\tiny
\caption{
Seven-category tutor-move taxonomy for controllable LLM tutoring, grounded in prior work on tutoring dialogue, classroom discourse, productive talk, and tutoring-strategy mining.
}
\label{tab:tutoring-move-taxonomy}
\resizebox{\textwidth}{!}{%
\begin{tblr}{
  colspec = {
    Q[c,m,0.025\textwidth]
    Q[l,m,0.09\textwidth]
    Q[l,m,0.55\textwidth]
    Q[l,m,0.27\textwidth]
  },
  row{1} = {font=\bfseries},
  hline{1,2,Z} = {1pt},
  hline{5,8} = {0.8pt},
  cell{2}{1} = {r=3}{bg=gray!10,c,m},
  cell{5}{1} = {r=3}{bg=gray!10,c,m},
  cell{8}{1} = {bg=gray!10,c,m},
}
\textbf{Type} & \textbf{Tutor Move} & \textbf{Definition} & \textbf{Example} \\

\rotlabel{Scaffolding} &
Feedback &
Evaluates student's answer without adding reasoning or follow-up \cite{Graesser2004,LIN2022194}. &
``That's correct.'' / ``You got the right operation.'' \\

&
Pump &
Invites broad continuation or elaboration without narrowing the response space \cite{Graesser2004,OCONNOR2019166}. &
``Go on.'' / ``What else?'' / ``Can you say more?'' \\

&
Focusing &
Asks the student to explain or justify part of their reasoning \cite{herbel2005questioning,doi:10.1177/0022487109339906,hagenah2018funneling,OCONNOR2019166}. &
``Why does that step work?'' / ``Where did you make that choice?'' \\

\rotlabel{Answer revealing} &
Hint/Funnelling &
Narrows solution space with a clue, reminder, correction, or guided direction without giving the answer \cite{Graesser2004,herbel2005questioning,hagenah2018funneling,LIN2022194,alic-etal-2022-computationally}. &
``What is the inverse operation here?'' / ``Look at the coefficient of $x$.'' \\

&
Prompt &
Nearly completes the step, leaving only a short word, value, symbol, or phrase for the student \cite{Graesser2004}. &
``So $12$ divided by $4$ is...?'' / ``So $x$ equals...?'' \\

&
Assertion &
Directly states the answer, fact, procedure, correction, or explanation \cite{Graesser2004}. &
``Divide both by $3$.'' / ``The answer is $x=4$.'' \\

\rotlabel{Reflect} &
Metacognition &
Asks the student to reflect on their understanding, confidence, confusion, strategy, or learning state \cite{OCONNOR2019166}. &
``How confident are you?'' / ``What made this confusing?'' \\

\end{tblr}%
}
\end{table*}

\subsection{Learning Tutoring Move Vectors}
\label{sec:learning-vectors}
To control tutoring moves at generation time, we learn move-specific residual-stream steering vectors across selected layers. For a target move $m$, frozen model $M$, tutoring context $q$, and selected residual-stream layers $\mathcal{S}$, PIVOT learns a multi-layer vector set $V^{(m)}=\{v_\ell^{(m)}\}_{\ell\in\mathcal{S}}$, where each layer-specific vector $v_\ell^{(m)}$ is added to the residual stream at its corresponding layer $\ell$. The selected layer set $\mathcal{S}$ determines where vectors are learned and applied, and can be varied for layer-ablation experiments. As shown in Figure~\ref{fig:method_fig}, learning follows an iterative generate--judge--optimise loop. Given a student--tutor context, the LLM generates candidate tutor responses with steering applied. A human-validated LLM judge labels each response by tutoring move, while relevance and fluency judges filter low-quality outputs. Target and non-target generations are then converted into preference pairs used to update the steering vectors before the next round.

\vspace{1mm} \noindent \textbf{Generate and Judge.}
At each round, we sample a tutoring context $q \sim \mathcal{X}$ and generate candidate tutor responses with the current vectors $V^{(m)}$. Intervention strength is controlled by scalars $\alpha_{\mathrm{prompt}}$ and $\alpha_{\mathrm{gen}}$, which scale steering at the final prompt token and during autoregressive generation, respectively. For online discovery, we use $\alpha_{\mathrm{prompt}}=0.1$ and $\alpha_{\mathrm{gen}}=1.0$, keeping prompt perturbations small while allowing steering directions to shape generation. We use three instruction types: (1) general tutoring instructions, (2) instructions targeting the current move, and (3) instructions targeting under-produced moves. The third type is updated online from judged outputs to maintain move diversity. See Appendix~\ref{appen:steer-model-prompt} for all prompts. Each response $y$ is evaluated by human-validated LLM judges (Section~\ref{sec:offline_evaluation}): $J_{\mathrm{move}}$ assigns move label $\hat{m}$, while $J_{\mathrm{rel}}$ and $J_{\mathrm{flu}}$ assign binary relevance and fluency labels. Relevant and fluent responses with $\hat{m}=m$ are added to the target pool, and responses with $\hat{m}\neq m$ to the non-target pool.

\vspace{1mm} \noindent \textbf{Preference Optimisation.}
We first construct online preference tuples $(q,y^+,y^-,c)$ from target and non-target pools, where $q$ is the tutoring context, $y^+$ is a validated target response, $y^-$ is a rejected non-target response, and $c=J_{\mathrm{move}}(q,y^-)$. Each tuple receives weight $w(q,y^+,y^-,c)=\max(1-\delta(y^+),\epsilon)w_c$, where $\delta(y^+)$ is an extra rule-based penalisation of malformed, or overly long responses, and $w_c$ is updated online to upweight frequent target confusions.
We then optimise the move-specific vectors across selected layers using weighted preference tuples. During optimisation, we set $\alpha_{\mathrm{prompt}}^{\mathrm{train}}=\alpha_{\mathrm{gen}}^{\mathrm{train}}=1.0$, to learn steering directions under normalised interventions. For each tuple, let
\[
\begin{aligned}
\Delta
=&\ 
\log \pi_{V^{(m)}}(y^+\mid q)
-\log \pi_{V^{(m)}}(y^-\mid q) \\
&-
\log \pi_0(y^+\mid q)
+\log \pi_0(y^-\mid q),
\end{aligned}
\]
where $\pi_{V^{(m)}}$ is the steered model and $\pi_0$ is the frozen unsteered model. We optimise
\[
\begin{aligned}
\mathcal{L}^{(m)}
=&\ 
\mathbb{E}_{(q,y^+,y^-,c)}
\left[
-w(q,y^+,y^-,c)
\log \sigma(\beta\Delta)
\right] \\
&+
\frac{\lambda}{|\mathcal{S}|}
\sum_{\ell\in\mathcal{S}}
\|v_\ell^{(m)}\|_2^2 .
\end{aligned}
\]
The optimisation updates all steering vectors jointly across layers. Algorithm~\ref{alg:pivot} summarises the procedure.

\begin{algorithm}[t]
\small
\SetAlgoNoLine
\caption{PIVOT: Online Preference-Based Steering Vector Learning}
\label{alg:pivot}

\KwIn{Contexts $\mathcal{X}$; target move $m$; frozen model $M$; judges $J$; layers $\mathcal{S}$}
\KwOut{Move vector set $V^{(m)}$}

Initialise $V^{(m)}$, preference set $\mathcal{D}$, rejected-class weights $w_c$

\For{$r = 1,\ldots,T$}{
    Generate candidate responses $y \sim M(x;V^{(m)})$ for $x \sim \mathcal{X}$
    \tcp*[r]{Self-discovery}

    Label candidates with $J_{\mathrm{move}}$, $J_{\mathrm{rel}}$, and $J_{\mathrm{flu}}$
    \tcp*[r]{Judging}

    Construct tuples $(x,y^+,y^-,c,w)$ and add them to $\mathcal{D}$
    \tcp*[r]{Online preferences}

    Update $w_c$ from rejected move confusions
    \tcp*[r]{Hard-negative weighting}

    Update $V^{(m)}$ by minimizing $\mathcal{L}^{(m)}$
    \tcp*[r]{Preference-vector update}
}

\end{algorithm}

\subsection{Inference-Time Steering}
\label{sec:inference-steering}
At inference time, PIVOT steers tutor responses by selecting one or more learned tutoring-move vectors, scaling their strength, injecting them into selected residual-stream layers, and combining them when multiple moves are requested. This produces a steered tutor response without updating the LLM.

For a target move $m$, PIVOT modifies the hidden state at each selected layer $\ell\in\mathcal{S}$ as
\[
h'_{\ell,t}=h_{\ell,t}+\gamma_t v_\ell^{(m)},
\]
where $\gamma_t=\alpha_{\mathrm{prompt}}$ for the final prompt token and $\gamma_t=\alpha_{\mathrm{gen}}$ for generated tokens. The prompt coefficient nudges the first generated token, while the generation coefficient is applied autoregressively and provides the main steering signal.

When multiple moves are selected, PIVOT combines their vectors at each layer:
\[
\tilde v_\ell=\sum_m \alpha_m w_\ell^{(m)}v_\ell^{(m)},
\]
where $\alpha_m$ is the move-specific steering strength and $w_\ell^{(m)}$ is an optional layer weight. The combined vector $\tilde v_\ell$ is then injected into the selected layers using the same token-dependent schedule.

% \subsection{Evaluation}
% We evaluated PIVOT on three tutoring datasets from different domains using expert annotation and a LLM-as-a-judge protocol as well as through a user study with teachers.

% % Datasets used to evaluate
% \subsubsection{Datasets.}

% % Offline evaluation method
% \subsubsection{Offline Evaluation}
% \label{sec:offline_evaluation}
% We evaluated generated tutor responses using the human-validated automatic judge described in Section~\ref{sec:judge}. The judge is prompted with the tutoring context, the generated tutor response, and the pedagogical move rubric. For each response, it evaluates alignment with the intended tutoring move, as well as relevance to the conversation history and fluency of the generated response.

\subsection{Evaluation}
\label{sec:evaluation}
We evaluated PIVOT with both offline automatic evaluation and a teacher user study.

\subsubsection{Datasets}
We evaluated PIVOT on three publicly available tutoring datasets from different domains. \textit{MathDial} contains 2,861 grade-school math tutoring dialogues between human teachers and LLM-simulated students with plausible misconceptions \cite{macina-etal-2023-mathdial}. \textit{ConvoLearn} contains 2,134 middle-school Earth Science dialogues based on 7th-grade questions, authored by K-12 teachers interacting with simulated students \cite{sharma2026convolearnlearningsciencesgrounded}.
\textit{StudyChat} contains real student interactions with an LLM chatbot in an undergraduate AI course. We filtered for conceptual and contextual questions, yielding 30,358 turns spanning programming, mathematics, assignment clarification, code explanation, and output interpretation \cite{10.1145/3785022.3785029}.
% We evaluated PIVOT on three tutoring datasets from different domains. \textit{MathDial} contains 2,861 one-to-one tutoring dialogues grounded in grade-school maths word problems, pairing human teachers with LLM-simulated students exhibiting plausible misconceptions \cite{macina-etal-2023-mathdial}. \textit{ConvoLearn} contains 2,134 middle-school Earth Science tutoring dialogues based on 7th-grade science questions, authored by K-12 teachers interacting with simulated students \cite{sharma2026convolearnlearningsciencesgrounded}. \cite{sharma2026convolearnlearningsciencesgrounded}. \textit{StudyChat} contains real-world student interactions with a LLM-chatbot in an undergraduate artificial intelligence course. We focus on conceptual questions about programming, Python libraries, computer science, tools, and mathematics, as these most closely resemble tutoring-oriented help-seeking behaviour \cite{10.1145/3785022.3785029}.

\subsubsection{Offline Evaluation}
\label{sec:offline_evaluation}
We evaluated tutor responses with three Qwen3.5-35B-A3B judges: tutor move hit rate, relevance, and fluency \cite{qwen3.5}. Each judge received the tutoring context, response, and rubric, with prompts provided in Appendix~\ref{appen:jude-prompts}. The tutor-move judge used the taxonomy in Table~\ref{tab:tutoring-move-taxonomy}, the relevance judge assessed whether the response was logically consistent with and relevant to the conversation history \cite{maurya-etal-2025-unifying}; and the fluency judge assessed grammar and readability. To validate the judges, two human annotators independently labelled 160 tutor responses across the three datasets. These responses were annotated with their surrounding dialogue context, comprising 327 total student and tutor turns. Annotators labelled each tutor response for move, relevance, and fluency, and we resolved disagreements to refine the rubrics. Human-human agreement exceeded $\kappa=0.88$ overall and for each tutor move, while human-LLM agreement exceeded $\kappa=0.89$; see Appendix~\ref{appen:judge-validation} for Cohen's $\kappa$ breakdown.

\subsubsection{User Study}
To evaluate whether PIVOT produced recognisable and pedagogically meaningful tutor behaviours, we conducted a user study with 30 teachers recruited via Prolific\footnote{\footnotesize \url{https://www.prolific.com}}, approved by the university ethics commission (Nr. [Anonymous]). Participants had primary or elementary mathematics teaching experience and completed demographic questions before the task (Appendix~\ref{appen:user-study}). The study focused on mathematics tutoring using three MathDial student scenarios: \textit{stuck or off-track}, \textit{partial answer}, and \textit{passive or unsure}.

After an introduction and practice round, participants used six teacher-facing tutor-move controls to steer a tutoring conversation for up to five turns using 0--100\% sliders (100\% corresponds to an $\alpha=1.5$): Broad guidance (\textit{pump}), Narrow guidance (\textit{focusing}), Hint (\textit{hint}), Direct explanation (\textit{assertion}), Feedback (\textit{feedback}), and Reflective question (\textit{metacognition}). We used Qwen3-30B-A3B-Instruct-2507 \cite{qwen3technicalreport} as the simulated student agent (prompt in Appendix~\ref{appen:user-study}). Participants iteratively adjusted controls and regenerated conversations until satisfied, then compared the final steered conversation against a neutral baseline generated from the same prompt with all controls disabled. Finally, participants completed a post-task questionnaire (see Appendix~\ref{appen:user-study}) covering controllability \cite{10.1145/3613904.3642462,davis1989technology}, pedagogical quality and teaching-intent alignment \cite{puech-etal-2025-towards,maurya-etal-2025-unifying}, usefulness \cite{davis1989technology}, preferred controls, and suggested improvements.

\section{Results}
We conducted a series of experiments examining the impact of steering strength $\alpha$, cross-model and cross-dataset vector transfer, and layer selection on the controllability–quality trade-off. We additionally conducted a user study to evaluate the usability and pedagogical relevance of the controls.
%We conducted four experiments to evaluate PIVOT. First, we swept steering strength $\alpha$ to measure the controllability-quality trade-off and tested whether learned directions could be subtracted to shape tutor-move behaviour. Second, we tested whether learned vectors generalised across datasets and transferred to a base model variant. Third, we ablated the layer structure to compare single-layer and multi-layer steering. Finally, we ran a teacher user study to evaluate whether the controls were usable and pedagogically meaningful.

\vspace{1mm} \noindent \textbf{Experimental protocol.}
All experiments used frozen Qwen3.5-9B \cite{qwen3.5}. Steering vectors were trained across every other layer from 6 to 28, covering early-to-middle through late layers, including the middle-to-late region found effective in prior Qwen steering work \cite{10.5555/3737916.3742333}, while using fewer intervention points than all-layer steering. 
Training used one four-GH200 node, with Qwen3.5-35B-A3B judges on a separate four-GH200 node, and learning all seven move vectors took about five wall-clock hours. We trained on $\mathcal{D}_{\mathrm{train}}$, built from MathDial and ConvoLearn with 2,810 conversations and 42,138 turns, and reserved StudyChat for evaluation only.
The main steering-strength set $\mathcal{D}{\mathrm{sweep}}$ contained 60 conversations, 20 per dataset, with 452 turns total. For compute-intensive ablations, we used a smaller $\mathcal{D}{\mathrm{ablate}}$ with 30 conversations, 10 per dataset, 282 turns, and conversations capped at 10 turns, after verifying that its controllability and quality trends matched $\mathcal{D}_{\mathrm{sweep}}$ (Appendix~\ref{appen:abl-testset}).

% The main steering-strength evaluation set $\mathcal{D}_{\mathrm{sweep}}$ consisted of 60 conversations with 20 from each dataset and 452 turns in total. For compute-intensive ablations, we a smaller evaluation dataset$\mathcal{D}_{\mathrm{ablate}}$ consisting of 30 conversations with 10 from each dataset, 282 turns, and conversations capped at 10 turns. We verified that the main controllability and quality trends on $\mathcal{D}_{\mathrm{ablat}}$ matched those on $\mathcal{D}_{\mathrm{sweep}}$ before using it for ablations (see Appendix \ref{appen:abl-testset}).

\subsection{PIVOT enables fine-grained steering}
\begin{figure}[t]
    \centering
    \includegraphics[width=0.9\columnwidth]{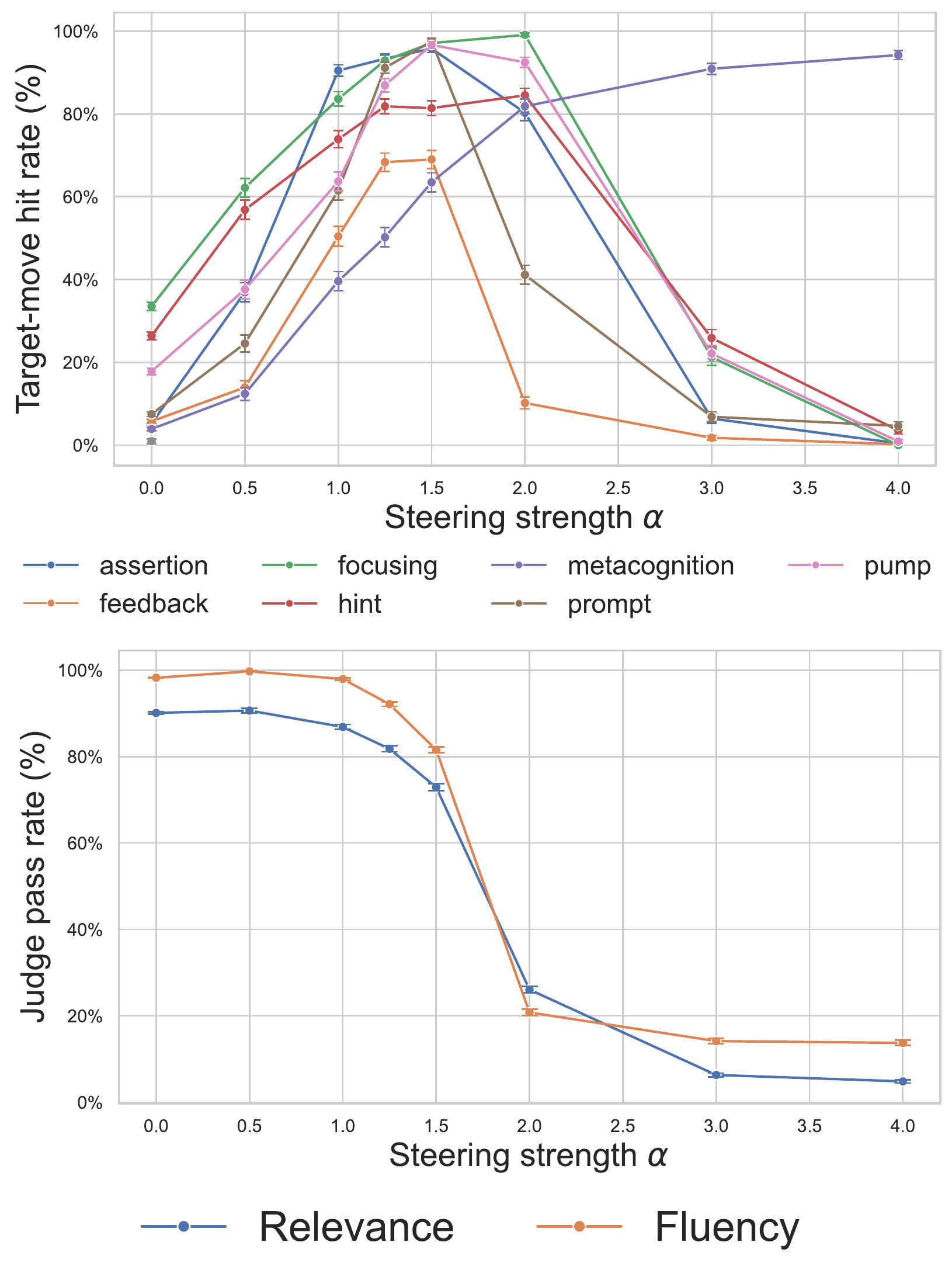}
    \caption{Steering-strength sweeps on $\mathcal{D}_{\mathrm{sweep}}$, showing target-move hit rate (top) and relevance and fluency pass rates (bottom).}
    \label{fig:alpha-sweep_quality_sweep}
\end{figure}

% Main alpha sweep on $\mathcal{D}_{\mathrm{sweep}}$
% Overall move controllability across steering strengths
% Best alpha / trade-off with quality
% Combining negative and postive 
% Alpha sweep, general results
We swept $\alpha$ on $\mathcal{D}_{\mathrm{sweep}}$ to assess its effect on tutor-move accuracy, relevance, and fluency. Figure~\ref{fig:alpha-sweep_quality_sweep} (top) shows target-move hit rates across steering strengths. At $\alpha=0$, we evaluated the unsteered model using the same tutoring prompt but without steering vectors, sampling five generations per context at temperature $0.8$ (prompt in Appendix~\ref{appen:steer-model-prompt}). For $\alpha>0$, steering vectors were added at the selected layers. Hit rates generally increased with $\alpha$, peaked at moderate strengths, and degraded at larger values, with the best range typically between $\alpha=1.0$ and $\alpha=2.0$. Figure~\ref{fig:alpha-sweep_quality_sweep} (bottom) shows corresponding relevance and fluency rates, which remained near baseline at moderate strengths but dropped sharply beyond $\alpha=2.0$. Appendix~\ref{appen:bipo-comparison} compares PIVOT with an offline BiPO-style baseline trained on the same preference pairs and optimisation budget; PIVOT achieves consistently higher tutor-move control.

We also evaluated vector subtraction and additive composition. As a case study, Feedback steering at $\alpha=1.0$ achieved a $48\%$ hit rate and was most frequently confused with Hint, Assertion, Focusing, and Pump (Appendix~\ref{appen:move-confusions}). Subtracting these vectors at $\alpha=-0.3$ increased the Feedback hit rate to $73\%$. We additionally tested additive move combinations, observing interpretable blended behaviours; for example, Assertion + Feedback produced responses that combined evaluative feedback with direct explanation. See Table~\ref{tab:additive-move-examples} (Appendix~\ref{appen:move-confusions}) for the detailed results.

% Figure on relevance and steering
% \begin{figure}[t]
%     \centering
%     \includegraphics[width=0.9\columnwidth]{figures/quality_sweep.pdf}
%     \caption{Relevance and fluency pass rates across steering strengths on $\mathcal{D}_{\mathrm{sweep}}$.}
%     \label{fig:quality-sweep}
%     \vspace{-0.5cm}
% \end{figure}

\vspace{-2mm}
\begin{tcolorbox}[
  colback=gray!10, 
  colframe=gray!50, 
  boxrule=0.5pt, 
  arc=2pt,
  fontupper=\small
]
PIVOT achieves best control-quality trade-off for $\alpha$ between $1.0$ and $2.0$, and learned directions can be added or subtracted to shape pedagogical behaviour.
\end{tcolorbox}

\subsection{PIVOT generalises across datasets and model variants}
Next, we investigated whether learned directions generalise across different datasets in $\mathcal{D}_{\mathrm{sweep}}$ and transfer across model variants. Table~\ref{tab:dataset-generalization} reports dataset-level results at $\alpha=1.5$: PIVOT achieved a hit rate of $88.9\%$ on held-out MathDial, $88.2\%$ on held-out ConvoLearn, and $80.6\%$ on StudyChat, which was unseen during training, suggesting transfer beyond the training conversations and to an unseen dataset. We also tested cross-variant transfer by applying Qwen3.5-9B steering vectors directly to Qwen3.5-9B-Base on $\mathcal{D}_{\mathrm{ablate}}$, since the base model shares the same architecture and pretraining, but lacks chat-oriented post-training that can support tutor-like responses. Table~\ref{tab:qwen9b-base-transfer} reports transfer results using $\alpha=1.5$ for all moves except Hint ($\alpha=1.0$) and Metacognition ($\alpha=2.5$), where those strengths performed better. The results show that learned directions can transfer to a closely related model variant without retraining and are not dependent on chat-oriented post-training.

% Table~\ref{tab:qwen9b-base-transfer} reports performance for the transfer setting. We used $\alpha=1.5$ for all moves except Hint, where $\alpha=1.0$ performed better, and Metacognition, where $\alpha=2.5$ performed better. The results suggest that transfer to a closely related model variant is possible without retraining, and that learned directions are not dependent on chat-oriented post-training.

% Figure on performance per data source
\begin{table}[t]
\centering
\small
\caption{Target-move hit rate across datasets $\mathcal{D}_{\mathrm{sweep}}$ at $\alpha=1.5$.}
\begin{tabular}{lcccc}
\toprule
Dataset & \makecell{Target-move\\Hit (\%)} & Relevance & Fluency \\
\midrule
ConvoLearn & $88.2 \pm 0.9$ & $96.2 \pm 0.7$ & $93.8 \pm 0.9$ \\
MathDial & $88.9 \pm 1.1$ & $75.0 \pm 2.0$ & $87.9 \pm 1.5$ \\
StudyChat & $80.6 \pm 1.2$ & $75.2 \pm 1.8$ & $86.8 \pm 1.4$ \\
\bottomrule
\end{tabular}

% Relevance and fluency exclude Feedback and Metacognition.
\label{tab:dataset-generalization}
\end{table}

% Figure on performance of Qwen-9b Base
% Generalisation on Qwen-9b Base
\begin{table}[t]
\centering
\small
\caption{Generalisation to Qwen-9B base on $\mathcal{D}_{\mathrm{ablate}}$.}
\resizebox{\columnwidth}{!}{%
\begin{tabular}{lccc}
\toprule
Move & \makecell{Target-move\\Hit (\%)} & Relevance & Fluency \\
\midrule
Assertion & $93.1 \pm 2.2$ & $90.8 \pm 2.6$ & $95.3 \pm 1.9$ \\
Feedback & $85.5 \pm 3.1$ & $68.3 \pm 4.2$ & $87.0 \pm 2.9$ \\
Focusing & $93.9 \pm 2.1$ & $89.8 \pm 2.7$ & $95.3 \pm 1.9$ \\
Hint & $64.1 \pm 4.2$ & $97.6 \pm 1.4$ & $98.5 \pm 1.1$ \\
Metacognition & $99.2 \pm 0.8$ & $99.2 \pm 0.8$ & $100.0 \pm 0.0$ \\
Prompt & $93.1 \pm 2.2$ & $90.3 \pm 2.7$ & $96.1 \pm 1.7$ \\
Pump & $93.9 \pm 2.1$ & $90.2 \pm 2.7$ & $94.6 \pm 2.0$ \\
\bottomrule
\end{tabular}}
\label{tab:qwen9b-base-transfer}
\end{table}

\begin{tcolorbox}[
  colback=gray!10, 
  colframe=gray!50, 
  boxrule=0.5pt, 
  arc=2pt,
  fontupper=\small
]
PIVOT generalises to unseen data; its learned directions are effective when transferred to a model variant.
\end{tcolorbox}

% PIVOT generalises to unseen data during vector learning, and its learned directions remain effective when transferred to a related model variant.

% \subsection{Multi-layer steering outperforms single-layer steering}
\subsection{Multi-layer outperforms single-layer} \label{sec:multilayer}
% 1-layer vs multi-layer
% Layer removal: remove early / middle / late -> appendix
% appendix: only early / only middle / only late in appendix

% Figure on single vs. multi-layer steering
% Steering strength is power-matched according to the norm of the learned steering vector. Single-layer is trained at layer 20; multi-layer uses $\alpha=1.5$. 
\begin{figure}[t]
    \centering
    \includegraphics[width=\columnwidth]{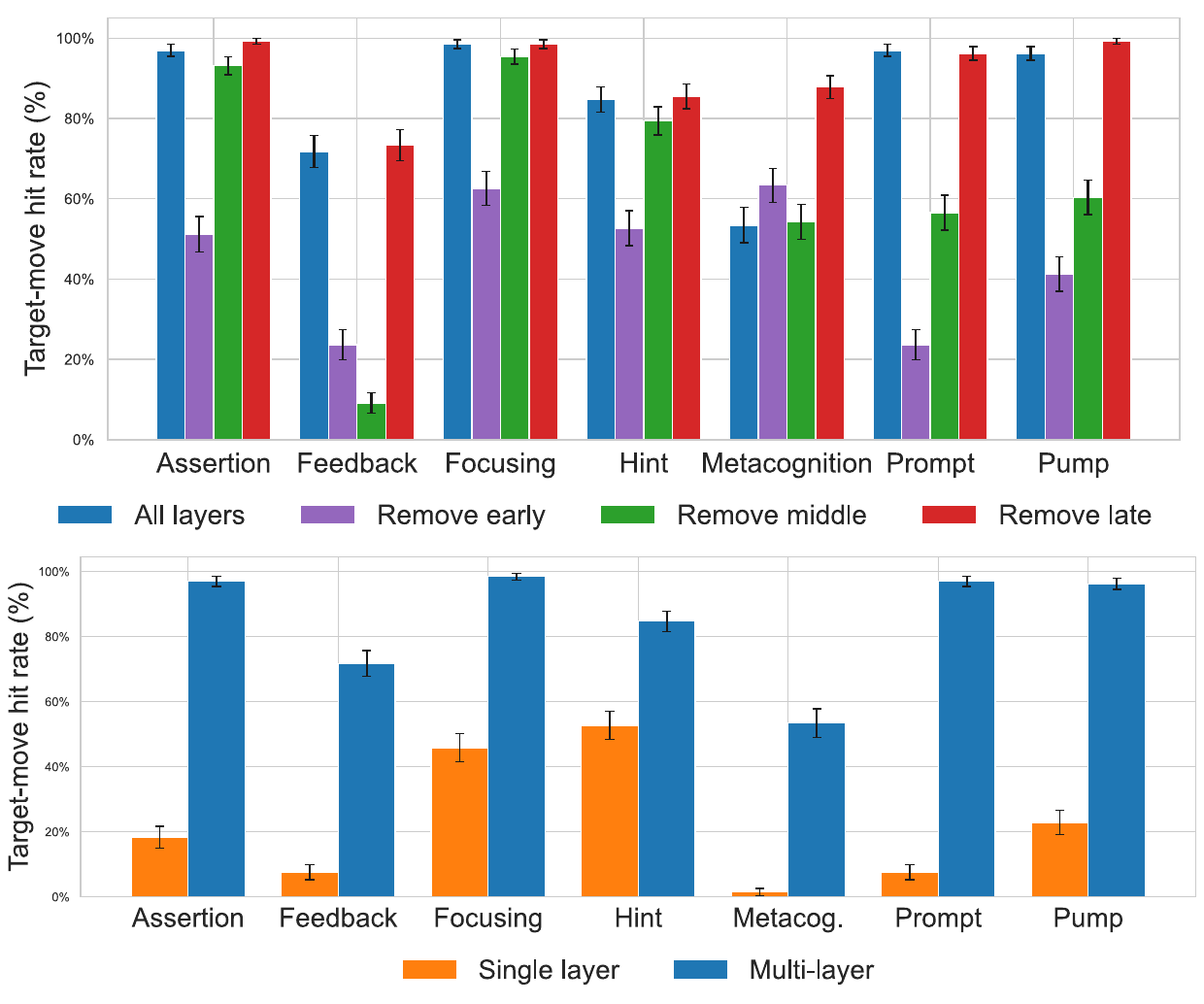}
    \caption{Layer-ablation analysis on $\mathcal{D}_{\mathrm{ablate}}$, showing leave-one-layer-group-out (top) and single-layer steering trained compared with multi-layer steering (bottom).}
    \label{fig:leave_one_out+single-vs-multi}
\end{figure}

% % Leave one out ablation for multi-layer steering
% \begin{figure}[t]
%     \centering
%     \includegraphics[width=\columnwidth]{figures/leaves_one_out_layers.pdf}
%     \caption{Leave-one-layer-group-out on $\mathcal{D}_{\mathrm{ablate}}$.}
%     \label{fig:leave-one-out-layers}
%     \vspace{-0.5cm}
% \end{figure}

% Single vs. multi-layer steerin
To test whether tutor-move steering benefits from multi-layer intervention, we compared our default setup, using every other residual-stream layer from 6 to 28, with a single-layer baseline at layer 20, a middle-to-late layer where prior Qwen steering work found effective interventions \cite{10.5555/3737916.3742333}. 
Figure \ref{fig:leave_one_out+single-vs-multi} (bottom) reports target-move hit rates for both settings. To control for steering power, we matched the single-layer baseline to the multi-layer condition by equating the total perturbation norm of the full vector set at $\alpha=1.5$, yielding a move-specific equivalent $\alpha$ for layer 20. Specifically, we scaled layer-20 vector by the ratio between the combined norm of the 12 multi-layer edits at $\alpha=1.5$ and the norm of the layer-20 vector, so that the single-layer intervention had comparable total injected-vector magnitude.
Across all tutor moves, multi-layer steering significantly outperformed the single-layer baseline, with paired permutation tests significant after Holm-Bonferroni correction ($p<.001$) for every move. This suggests that reliable tutor-move control benefits from representing each move vector across multiple residual-stream layers rather than using a single layer-specific intervention.
To identify which layers within our multi-layer range (6–28) carried the most signal, we performed post-hoc layer-group ablations: after training the full vectors once, we removed early (6, 8, 10, 12), middle (14, 16, 18, 20), or late (22, 24, 26, 28) layers at inference time.
Figure \ref{fig:leave_one_out+single-vs-multi} (top) compares each removal to the all-layer setting using paired permutation tests with Holm-Bonferroni correction. Removing early layers significantly reduces hit rates for six of seven moves, except Metacognition; removing middle layers selectively hurts Feedback, Prompt, and Pump; and removing late layers does not reduce any move and improves Metacognition. Thus, early layers are most important, middle layers support some moves, and late layers are largely unnecessary, consistent with prior work showing behavioural representations can emerge before final layers and steering effects can peak earlier \cite{rimsky-etal-2024-steering}. Additional isolated-layer experiments are in Appendix~\ref{appn:layer-groups}.

\begin{tcolorbox}[
  colback=gray!10, 
  colframe=gray!50, 
  boxrule=0.5pt, 
  arc=2pt,
  fontupper=\small
]
Multi-layer steering significantly outperforms single-layer steering for all tutor moves, with early-to-middle layers contributing most to reliable control.
\end{tcolorbox}

% and use controls meaningfully
% \subsection{Teachers prefer PIVOT-steered tutor conversations}
\subsection{Teachers prefer PIVOT-steered responses}
\label{sec:teacher-facing-study}

% Figure showing Likert-scale ratings of interface
\begin{figure}[t]
    \centering
    \includegraphics[width=\columnwidth]{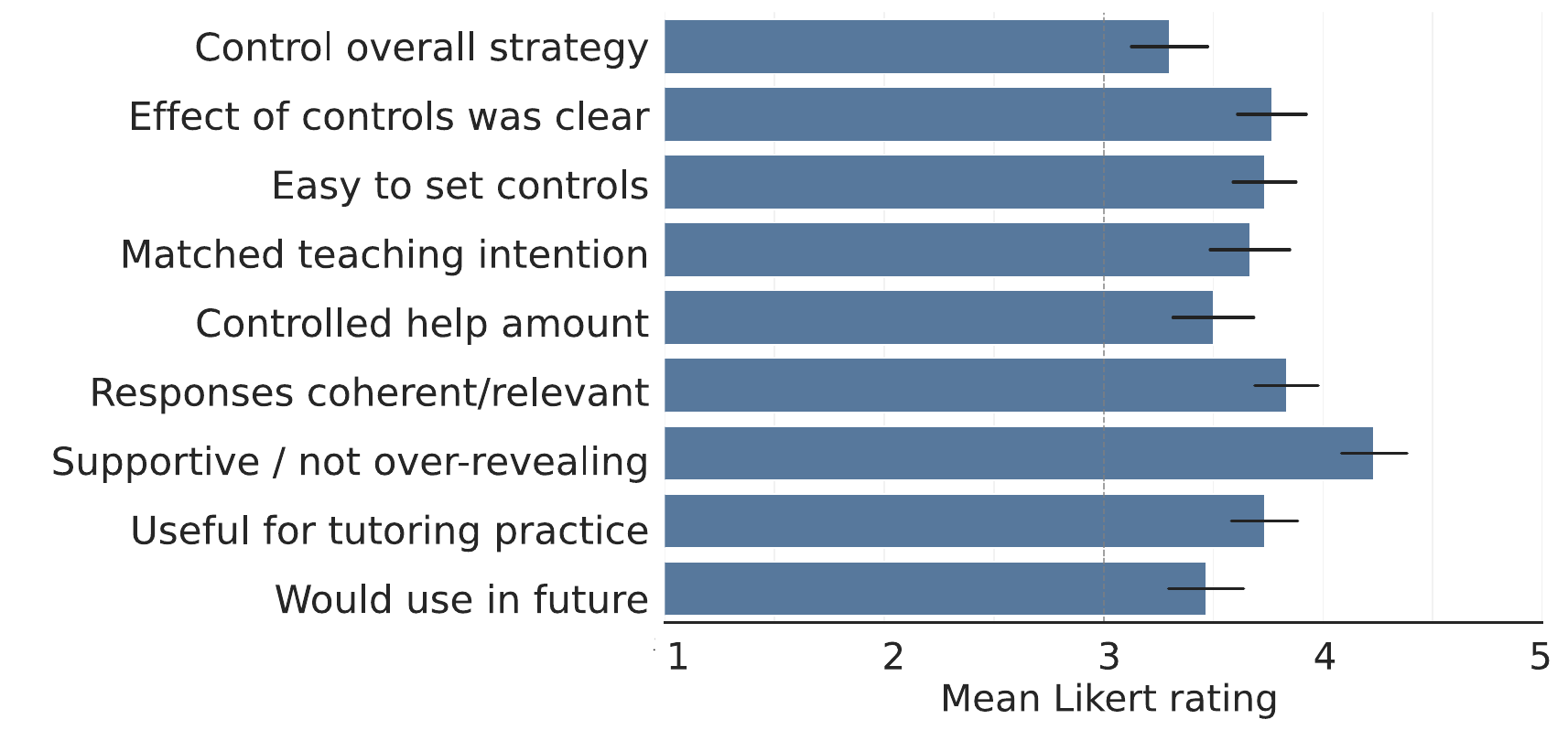}
    \caption{Teacher ratings of the controllable tutoring interface. Ratings are on a 1--5 Likert scale.}
    \label{fig:teacher-ratings}
\end{figure}

% Figure showing move usage by sceanrio
\begin{figure}[t]
    \centering
    \includegraphics[width=\columnwidth]{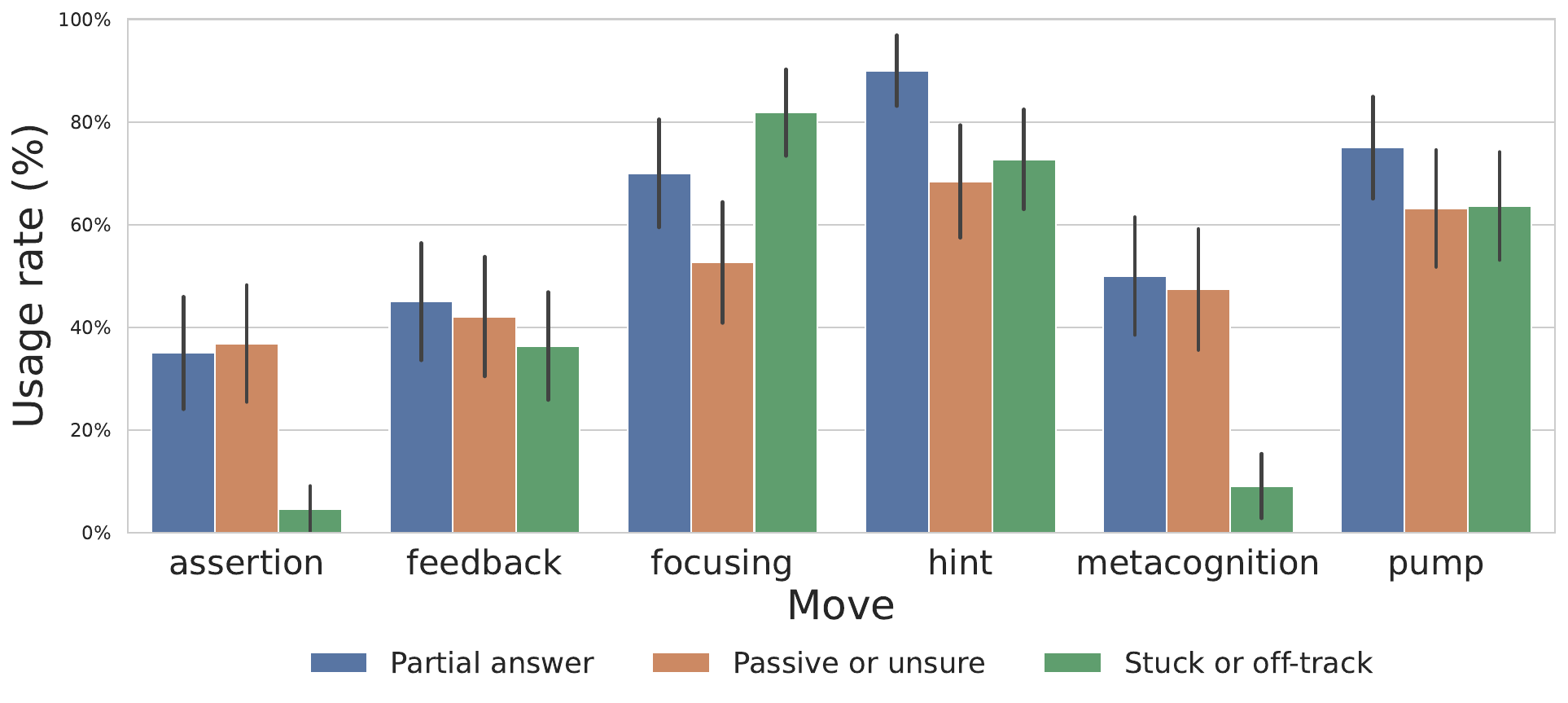}
    \caption{Move usage by instructional scenario. Teachers selected different controls depending on the state.}
    \label{fig:move-usage-scenario}
\end{figure}

In our user study, we tested whether learned steering vectors could guide tutor responses through teacher-facing controls. Teachers rated the controls positively on a 1-5 Likert scale (Figure~\ref{fig:teacher-ratings}), for producing supportive responses without over-revealing ($mean=4.23$), coherence and relevance ($mean=3.83$), and clear control effects ($mean=3.77$). Ease of use, teaching-intent match, usefulness, and future use were rated positively, suggesting that teachers found the controls interpretable and useful.

Most teachers preferred the controlled conversation over a neutral one generated with the same prompt but no steering: 22/30 participants (73.3\%) preferred the controlled setting, compared with 3/30 (10.0\%) preferring neutral and 5/30 (16.7\%) reporting no preference. Open-ended responses suggest that teachers valued guidance without over-direction: one described the controlled conversation as ``more guided whilst allowing the student to use their current knowledge,'' while another said it ``encouraged the student to expand on their thinking.’’ Participants particularly valued Hint and Focusing controls because they nudged the student toward ``figuring out'' the answer and supported gradual progress without sounding ``dismissive.’’

Finally, teachers used different moves across scenarios (Figure~\ref{fig:move-usage-scenario}). Hint was common throughout, while Focusing was especially frequent for stuck or off-track students. Partial-answer scenarios showed high use of Hint, Pump, and Focusing, suggesting that teachers combined targeted guidance with opportunities for reasoning. Passive or unsure scenarios led to relatively more Assertion, alongside continued Hint and Pump use. These patterns suggest that teachers treated the controls as student-state-responsive pedagogical actions rather than arbitrary generation knobs. Refer to Appendix~\ref{appen:user-study} for additional move-combination and scenario-specific analyses.

% Additional move-combination and scenario-specific analyses are in Appendix~\ref{appen:user-study}.

\begin{tcolorbox}[
  colback=gray!10, 
  colframe=gray!50, 
  boxrule=0.5pt, 
  arc=2pt,
  fontupper=\small
]
Teachers preferred steered conversations and found the controls clear, usable, and pedagogically useful.
\end{tcolorbox}

\section{Conclusion}
We introduced PIVOT, a framework for inference-time activation steering of pedagogical tutor moves. PIVOT turns tutor behaviours such as feedback, pumping, focusing, hinting, prompting, assertion, and metacognitive reflection into steerable residual-stream directions. It combines a seven-category taxonomy with online judge-driven preference construction, using target and confusable non-target generations to optimise multi-layer steering vectors. Our experiments showed that PIVOT improves tutor-move control while preserving relevance and fluency, transfers to unseen tutoring datasets, and benefits from coordinated interventions across multiple layers. We further showed that the learned directions can be subtracted or added to shape tutor behaviour. Finally, a teacher user study showed that educators can meaningfully use the controls and generally prefer steered conversations. These results suggest that activation steering is a promising mechanism for giving educators finer-grained control over how LLM tutors respond, without retraining model weights or relying solely on prompts.

\section{Limitations}
PIVOT improves inference-time control over tutor moves, but several limitations remain. First, our evaluation focuses on tutor-response quality and teacher preferences rather than direct student learning outcomes. Although teachers preferred controlled tutor conversations and rated the controls as pedagogically useful, future work should measure how interaction with PIVOT-steered tutors affects student learning, persistence, and performance over time. This could be done in a classroom setting. Second, tutor-move controls are set once for the full conversation, so teachers must manually intervene and adjust the steering if the desired pedagogical move changes over time. Future work should develop a dynamic steering-vector scheduler that adapts tutor moves automatically based on student progress throughout the conversation. Finally, our user study focuses on mathematics tutoring with a simulated student agent, so future work should focus on examining whether the results could fully generalise to real students, longer instructional settings, or other subject areas.

Transfer to Qwen-9B-Base is strong overall, but varies by move, with Hint showing the largest remaining gap. This is expected under cross-variant transfer, since hinting requires a fine balance between useful guidance and answer disclosure. Future work could study move-specific calibration when transferring vectors to other model variants.

\section{Ethics Statement and Potential Risks}
PIVOT is designed to support controllable LLM tutoring, not to replace human educators. Because tutor responses can affect student understanding, confidence, and problem-solving behaviour, any deployment should include teacher oversight, transparent controls, and safeguards against over-reliance on generated feedback. Our teacher study was approved by the university ethics committee, and participants provided informed consent, were recruited according to the stated eligibility criteria, and were compensated for their time.

The main risks of PIVOT come from misuse or over-interpretation of the controls. Steering vectors can make tutor behaviour more controllable, but they do not guarantee pedagogical correctness, improved learning outcomes, or safety in all contexts. Overly strong or poorly chosen controls may over-direct students, reveal answers too early, or produce responses that appear pedagogically appropriate without actually supporting learning. Our evaluation therefore measures tutor-move control, relevance, fluency, and teacher preferences, but should not be interpreted as evidence of improved student learning without future studies with real learners and learning outcomes. 

We release code, prompts, and annotations to support reproducibility, but these resources should be used with their documented limitations. Systems using PIVOT should report the intended educational setting, monitor generated tutor responses, and evaluate possible failure modes before use with students.
\section{Acknowledgements}
We acknowledge that the use of AI assistants (ChatGPT) was limited to polishing the language of the original paper. It was used for proofreading and refining grammar, spelling, and phrasing.

\bibliography{custom}
\newpage
\appendix
\section{Tutor Generation Prompt}
\label{appen:steer-model-prompt}

We use the following prompt to generate tutor responses for the move-balanced generation setting. The prompt instructs the tutor model to produce a single concise utterance, follow exactly one tutoring move, and avoid mixing pedagogical strategies within the same response. During preference construction, we append an additional move-bias instruction to the base prompt. We use the target-move instruction to elicit positive candidates and the under-produced-move instruction to encourage generations from moves that are currently underrepresented in the preference pool.

\newtcblisting{bluepromptbox}[1]{
    title=\textbf{#1},
    colback=blue!4,
    colframe=blue!25,
    coltitle=black,
    fonttitle=\bfseries,
    boxrule=0.5pt,
    arc=1mm,
    left=6pt,
    right=6pt,
    top=5pt,
    bottom=5pt,
    breakable,
    listing only,
    listing options={
        basicstyle=\ttfamily\footnotesize,
        breaklines=true,
        breakatwhitespace=true,
        columns=fullflexible,
        keepspaces=true,
        showstringspaces=false
    }
}

\begin{bluepromptbox}{Move-Biased Generation Instruction}
{base_tutor_generation_prompt}

For this one response, USE the move: {target_move}
If it truly cannot fit, choose the closest single move, but do not blend moves.
No praise, no agreement, no feedback unless the chosen move is Feedback.
Do not mention the move label.
\end{bluepromptbox}

\begin{bluepromptbox}{Under-Produced Move Generation Instruction}
{base_tutor_generation_prompt}

For this one response, USE the move: {under_produced_move}
If it truly cannot fit, choose the closest single move, but do not blend moves.
No praise, no agreement, no feedback unless the chosen move is Feedback.
Do not mention the move label.
\end{bluepromptbox}
Here, \texttt{\{move\_name\}} is instantiated as either the target move or an under-produced move during preference construction.

\begin{bluepromptbox}{Tutor Generation Prompt}
You are a tutor. Continue the dialogue with exactly one brief, targeted tutor reply.
Choose EXACTLY ONE tutoring move from this set: Feedback, Pump, Focusing, Hint, Prompt, Assertion, Metacognition, or Generic.
Follow the move definitions strictly. Do not mix moves in a single utterance.
If the user message explicitly says to USE a particular move for this one response, you MUST obey that move.
Be concise and critical: do not default to agreement, praise, or encouragement fillers.
Avoid praise-y or agreement openers such as "I'm so glad", "That's fantastic", "Great job", "That's a great start", "That sounds right", or "Yes, exactly" unless the chosen move is Feedback and the whole utterance is only that evaluation.
Return only the tutor utterance. Do not mention the move name. Do not think.
Only Use ENGLISH. No Other Languages. Only English.

BALANCED DEFINITIONS

- Pump: ask an open "what else" continuation question that does NOT point to a specific missing condition, quantity, or subgoal. Do NOT say things like "what other information from the problem statement...", "what do you need to find/calculate...", or "what is the next step...". Provide NO hint, NO clue, and NO new information. Do not evaluate (no praise, no agreement, no feedback). Examples:  "What else could be going on here?" / "What else do you notice?"

- Focusing: question the student's line of thinking by directly targeting what they said or did. Ask them to justify/clarify a specific step they already made. Provide NO new information. Examples: "How did you decide that?" / "Why did you do that step?"

- Hint (Hint/Funneling): provide ONE new clue or piece of information that helps without giving away the answer. The student still has a lot to do. Examples: "Remember this is only half of what you need." / "Compare what happens when X is bigger versus smaller."

- Prompt: give almost the full answer/solution setup and ask for a VERY short completion (one word/number/short phrase) to finish it. Do not ask a broad question. Do not add a new clue. Examples: "So the total is ... ?" / "That means x = ... ?"

- Assertion: give the full final answer or full correction such that NOTHING substantive is needed from the student anymore. No real follow-up question. Examples: "The answer is 24." / "A thinner atmosphere traps less heat, so it gets colder."

- Feedback: only judge correctness (right/wrong/partly right) and stop. No hint, no clue, no next-step question. Examples: "That's right." / "Not quite."

- Metacognition: ask about the student's FEELINGS or learning experience, NOT the task (confidence, confusion, progress). High-level reflection about the course/concepts rather than this problem step. Examples: "How confident do you feel about this topic?" / "What part feels most confusing overall?"

- Generic: generic/unrelated text, off-task. Examples: "What's your favorite color?" / "How old are you?"
\end{bluepromptbox}
\newtcblisting{promptbox}[1]{
    title=\textbf{#1},
    colback=gray!5,
    colframe=gray!45,
    coltitle=black,
    fonttitle=\bfseries,
    boxrule=0.5pt,
    arc=1mm,
    left=6pt,
    right=6pt,
    top=5pt,
    bottom=5pt,
    breakable,
    listing only,
    listing options={
        basicstyle=\ttfamily\footnotesize,
        breaklines=true,
        breakatwhitespace=true,
        columns=fullflexible,
        keepspaces=true,
        showstringspaces=false
    }
}

\section{Judge Prompts}
\label{appen:jude-prompts}

We use three LLM-as-a-judge prompts to evaluate the final tutor utterance in each student--tutor conversation. The relevance judge measures whether the response remains grounded in the tutoring context, the fluency judge checks only surface language quality, and the tutor-move judge assigns the response to one of the tutoring-move categories. All judges use the full conversation history but label only the final tutor utterance.

\begin{promptbox}{Relevance Judge Prompt}
You are a strict evaluator of tutor response quality.
Your job is to label the FINAL tutor utterance for RELEVANCE to the tutoring conversation.

You must use the full conversation history to decide whether the final tutor utterance belongs in this conversation. Judge only the final tutor utterance, but compare it against the full history.

Relevance means: the final tutor utterance stays on the same problem, concept, student reasoning, or instructional goal as the conversation. It may respond to the student's most recent turn, or it may appropriately refer back to something earlier in the conversation to scaffold the student. It should be labeled not relevant only when it derails to an unrelated topic, answers a different problem, or is generic/random in a way that ignores the tutoring context.

Binary label:
1 = Relevant: on topic for the current tutoring conversation, including appropriate references to earlier turns.
0 = Not relevant: unrelated, derailed, generic in a way that ignores the conversation, or responding to a different topic/problem.

Output a JSON object matching the required schema, with a short explanation.
\end{promptbox}

\begin{promptbox}{Fluency Judge Prompt}
You are a strict evaluator of tutor response quality.
Your job is to label the FINAL tutor utterance for SURFACE LANGUAGE FLUENCY ONLY.

Fluency means only that the final tutor utterance is grammatical, readable, and linguistically well-formed as a standalone piece of text. Check for spelling, missing words, broken syntax, repeated fragments, excessive punctuation, malformed phrases, garbled wording, unfinished output, special tokens, or oversteering artifacts.

Do not judge logical flow, topical relevance, usefulness, correctness, teaching quality, or whether the sentence follows naturally from the conversation. A line can be fluent even if it is irrelevant, unhelpful, or logically odd in context.

Binary label:
1 = Fluent: grammatical, readable, complete, and understandable as text.
0 = Not fluent: ungrammatical, garbled, repetitive, unfinished, malformed, full of excessive punctuation, special tokens, or broken wording.

Examples of not fluent:
- "Can you can you can you what the the"
- "<|im_start|> teacher teacher"
- "Think about the the the because maybe ??"
- An unfinished sentence that cannot be understood.

Examples of fluent:
- "What specific evidence from the Jurassic Park movies connects dinosaurs to the movement of continents?"
- "Can you explain why you chose that step?"
- "That is close, but check the number you multiplied by."

Output a JSON object matching the required schema, with a short explanation.
\end{promptbox}

\begin{promptbox}{Tutor-Move Judge Prompt}
Classify only the final tutor utterance, but use the full chat history.
Your first job is to decide whether the final tutor utterance adds new information relative to the previous turns, or only confirms/restates what the student already said.
Choose exactly one label and return only JSON.

History rule:
- Read the chat history before labeling. Do not classify the final utterance in isolation.
- If the tutor repeats, paraphrases, accepts, or lightly rewords an idea already established by the student, that is not new information.
- If the tutor gives a new fact, value, mechanism, analogy, relation, missing operation, or explicit scaffold, that is new information.
- If there is a substantive follow-up question, classify the function of that question/scaffold unless the tutor has fully supplied the answer and the question is only a check like "do you understand?"

Labels:
0 Feedback/Confirmation/Rejection:
  Confirms, rejects, or evaluates the student's academic content. Can include a short restatement/paraphrase of what the student already established. No new information and no substantive follow-up.

1 Pump:
  On-task broad continuation. The tutor asks for broad elaboration, another idea, a whole-solution walkthrough, a broad causal/impact/consequence explanation, a meaning/definition, or what led to a content conclusion without adding any new clue, constraint, criterion, or direction that narrows the answer space. It shouldn't be a completion, that's a prompt. It does not add a new scaffold.
  Important: a broad "what else / what other" continuation question is Pump ONLY when it does not point to a specific missing condition, quantity, or subgoal (e.g., "What other piece of information from the problem statement haven't you used yet?" is usually Hint/Funneling because it directs the student to search for a missing condition).

2 Focusing:
  "Go back to your own step." The tutor explicitly points to the student's own idea, answer, value, step, decomposition, or named concept and asks the student to inspect, explain, justify, calculate from, or reason from it. No new help is added. Do not use Focusing for a general next-consequence question unless the question clearly targets the student's own stated mechanism or step. If the tutor anchors to one specific concept the student already mentioned and asks how it fits, works, or matters in the larger process, that is Focusing rather than Pump.
  If the tutor asks a pure clarification question like "What does X mean?" and DOES NOT provide the definition/explanation itself, that is Focusing when X is something the student just used.

3 Hint/Funneling:
  "Here is something new to help." The tutor adds a new clue, fact, mechanism, relation, analogy, missing operation, explicit scaffold, or targeted criterion that narrows the answer space toward one intended answer or concept, while still leaving substantive work to the student. If only a single short completion remains after the tutor's setup, use Prompt instead.

4 Prompt:
  The tutor asks for the next short value, word, symbol, or calculation result on an already active step-by-step path. In Math this includes the next required arithmetic result after the relevant relation or quantities have been stated. In non-math tutoring, this also includes asking for a single named fact or short completion after the tutor has already narrowed the target enough that only a one-word or very short phrase answer remains. Prompt is about response shape: the student should only need a very short completion.
  Do not use Prompt for broad "what else / what other items / what else do you need to find" questions that are not asking for a single short completion.

5 Assertion:
  The tutor directly supplies the needed final or intermediate answer, value, formula, fact, or explanation. No substantive extra information is required from the student in that turn. If the tutor already gave the answer and only adds a light check like 'okay?' or 'does that make sense?', still use Assertion.

6 Reflection/Metacognition:
  Learning-process self-report about feelings, confidence, confusion, challenge, ease, progress, or learning experience. If the question is about usefulness, applications, what else comes to mind, evidence, source, reasoning, or how the student reached a content conclusion, use Pump or Focusing instead.

7 Generic:
  Social, affective, not reflective, metacognitive or administrative utterance that does not evaluate academic correctness and does not teach the content.

Priority rules:
1. Social/affective closure with no academic evaluation -> Generic.
2. Off-task learning-process self-report -> Reflection/Metacognition.
3. No substantive follow-up question:
   - Restates/evaluates what student already said -> Feedback.
   - Supplies new needed content/value/explanation -> Assertion.
4. Substantive follow-up question:
   - If the expected student response is only a short completion already set up by the tutor, including a single factual word or short phrase in conceptual tutoring -> Prompt.
   - Tutor added new scaffold before/inside the question and the student still must reason from it -> Hint/Funneling.
   - Tutor explicitly points only to student's own step/idea -> Focusing.
   - Broad on-task elaboration, whole-solution reasoning, meaning/definition, or next-consequence/impact question not explicitly tied to the student's own reasoning -> Pump.

Dataset contexts: {dataset_contexts}
Question: {question}
Conversation history: {history}
Final tutor utterance: {final_utterance}
\end{promptbox}

\section{Judge Validation}
\label{appen:judge-validation}

We validated the tutor-move, relevance, and fluency judges against human annotations. For tutor moves, Tables~\ref{tab:judge-validation-datasets} and~\ref{tab:judge-validation-categories} report human--human agreement overall/by dataset and by category, while Tables~\ref{tab:judge-validation-llm-datasets} and~\ref{tab:judge-validation-llm-categories} report human--LLM agreement. For relevance and fluency, both human--human and human--LLM agreement reached 100\%, suggesting that these binary surface-quality and context-fit judgements were comparatively simple for the LLM judge.

\begin{table}[t]
\centering
\caption{Human--human agreement for tutor-move annotations overall and by dataset.}
\label{tab:judge-validation-datasets}
\small
\setlength{\tabcolsep}{8pt}
\renewcommand{\arraystretch}{1.12}
\begin{tabular}{lc}
\toprule
\textbf{Dataset} & \textbf{Cohen's $\kappa$} \\
\midrule
Overall & $0.929$ \\
ConvoLearn & $0.906$ \\
MathDial & $0.939$ \\
StudyChat & $1.000$ \\
\bottomrule
\end{tabular}
\end{table}

\begin{table}[t]
\centering
\caption{Human--human agreement using Cohen's $\kappa$ by tutor-move category.}
\label{tab:judge-validation-categories}
\small
\setlength{\tabcolsep}{8pt}
\renewcommand{\arraystretch}{1.12}
\begin{tabular}{lc}
\toprule
\textbf{Category} & \textbf{Cohen's $\kappa$} \\
\midrule
Feedback & $0.930$ \\
Pump & $0.922$ \\
Focusing & $0.846$ \\
Hint/Funneling & $0.916$ \\
Prompt & $0.956$ \\
Assertion & $0.982$ \\
Metacognition & $0.886$ \\
Generic & $1.000$ \\
\bottomrule
\end{tabular}
\end{table}

\begin{table}[t]
\centering
\caption{Human--LLM agreement for tutor-move annotations overall and by dataset.}
\label{tab:judge-validation-llm-datasets}
\small
\setlength{\tabcolsep}{8pt}
\renewcommand{\arraystretch}{1.12}
\begin{tabular}{lc}
\toprule
\textbf{Dataset} & \textbf{Cohen's $\kappa$} \\
\midrule
Overall & $0.954$ \\
ConvoLearn & $1.000$ \\
MathDial & $1.000$ \\
StudyChat & $0.906$ \\
\bottomrule
\end{tabular}
\end{table}

\begin{table}[t]
\centering
\caption{Human--LLM agreement using Cohen's $\kappa$ by tutor-move category.}
\label{tab:judge-validation-llm-categories}
\small
\setlength{\tabcolsep}{8pt}
\renewcommand{\arraystretch}{1.12}
\begin{tabular}{lc}
\toprule
\textbf{Category} & \textbf{Cohen's $\kappa$} \\
\midrule
Feedback & $1.000$ \\
Pump & $1.000$ \\
Focusing & $1.000$ \\
Hint/Funneling & $0.890$ \\
Prompt & $1.000$ \\
Assertion & $0.936$ \\
Reflection/Metacognition & 1.000 \\
Generic & $1.000$ \\
\bottomrule
\end{tabular}
\end{table}
\section{Isolated Layer-Group Steering}
\label{appn:layer-groups}

In addition to the leave-one-layer-group-out ablation in Section~\ref{sec:multilayer}, we trained steering vectors using each layer group in isolation and compared them with the full all-layer setting. Specifically, we trained separate vector sets using only early layers $(6,8,10,12)$, only middle layers $(14,16,18,20)$, or only late layers $(22,24,26,28)$, and evaluated each setting on $\mathcal{D}_{\mathrm{ablate}}$. This experiment tests whether the performance drop in the removal study is due to a layer group being useful on its own, or to its contribution as part of the full multi-layer intervention.

\begin{figure}[t]

    \centering

    \includegraphics[width=\columnwidth]{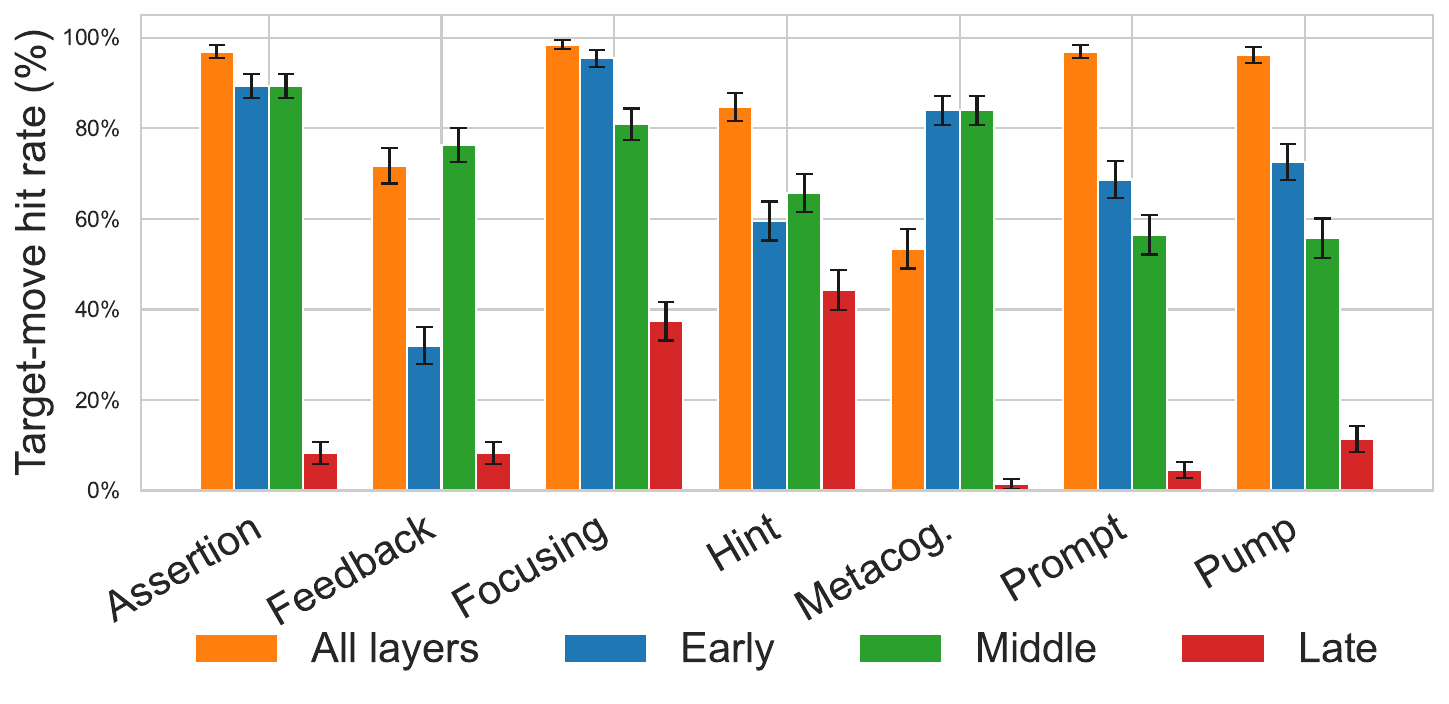}

    \caption{Target-move hit rates when training steering vectors on all selected layers, only early layers, only middle layers, or only late layers on $\mathcal{D}_{\mathrm{ablate}}$.}

    \label{fig:isolated-layer-groups}

\end{figure}

As shown in Figure~\ref{fig:isolated-layer-groups}, all-layer steering generally gives the strongest and most consistent control. Early-layer steering remains competitive for several moves, especially Assertion, Focusing, Metacognition, and Pump, while middle-layer steering performs well for Feedback and Metacognition. Late-layer steering performs substantially worse for most moves, suggesting that late residual-stream interventions alone are insufficient for reliable tutor-move control. These results are consistent with the leave-one-group-out ablation: early and middle layers carry most of the useful steering signal, while late layers contribute less to robust move control.

\section{Dynamic Weighting Ablation}
\label{appen:dynamic-weighting}

We ablated the online dynamic weighting used during preference-vector learning by training two variants: one with fixed equal negative-class weights and one with dynamic weights updated from the observed confusion distribution. The dynamic version upweights non-target moves that are frequently confused with the target move, encouraging the optimiser to focus on harder contrasts.

\begin{figure}[t]
    \centering
    \includegraphics[width=\columnwidth]{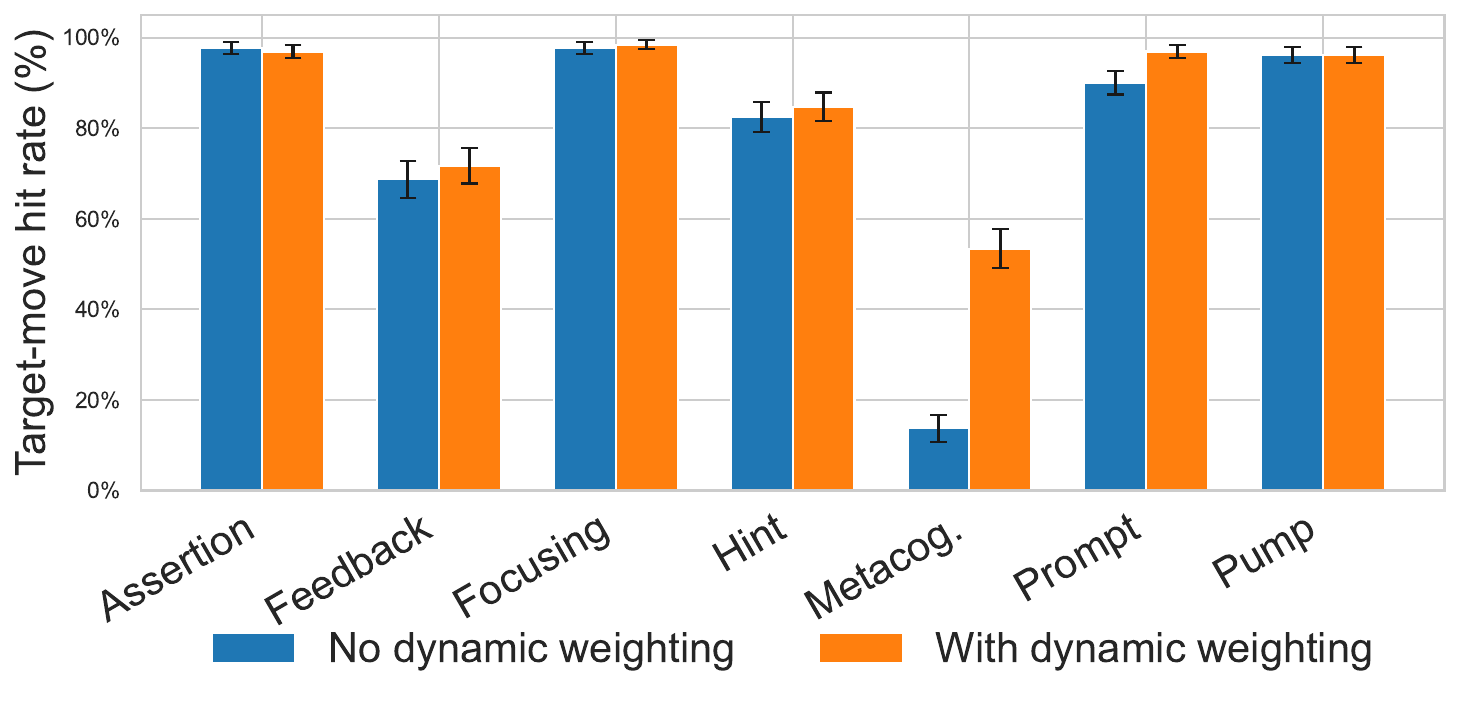}
    \caption{Target-move hit rates with and without dynamic negative-class weighting during steering-vector training.}
    \label{fig:dynamic-weighting-ablation}
\end{figure}

Figure~\ref{fig:dynamic-weighting-ablation} shows that dynamic weighting slightly improves target-move control for most tutor moves. The effect is especially large for Metacognition, where the hit rate increases substantially, and is also visible for Feedback, Hint, and Prompt. Assertion, Focusing, and Pump remain strong in both settings, suggesting that dynamic weighting is most useful for moves with harder or more imbalanced confusions.

\section{Move Confusions Across Steering Strengths}
\label{appen:move-confusions}

\begin{figure*}[t]
    \centering
    \begin{subfigure}[t]{0.48\textwidth}
        \centering
        \includegraphics[width=\linewidth]{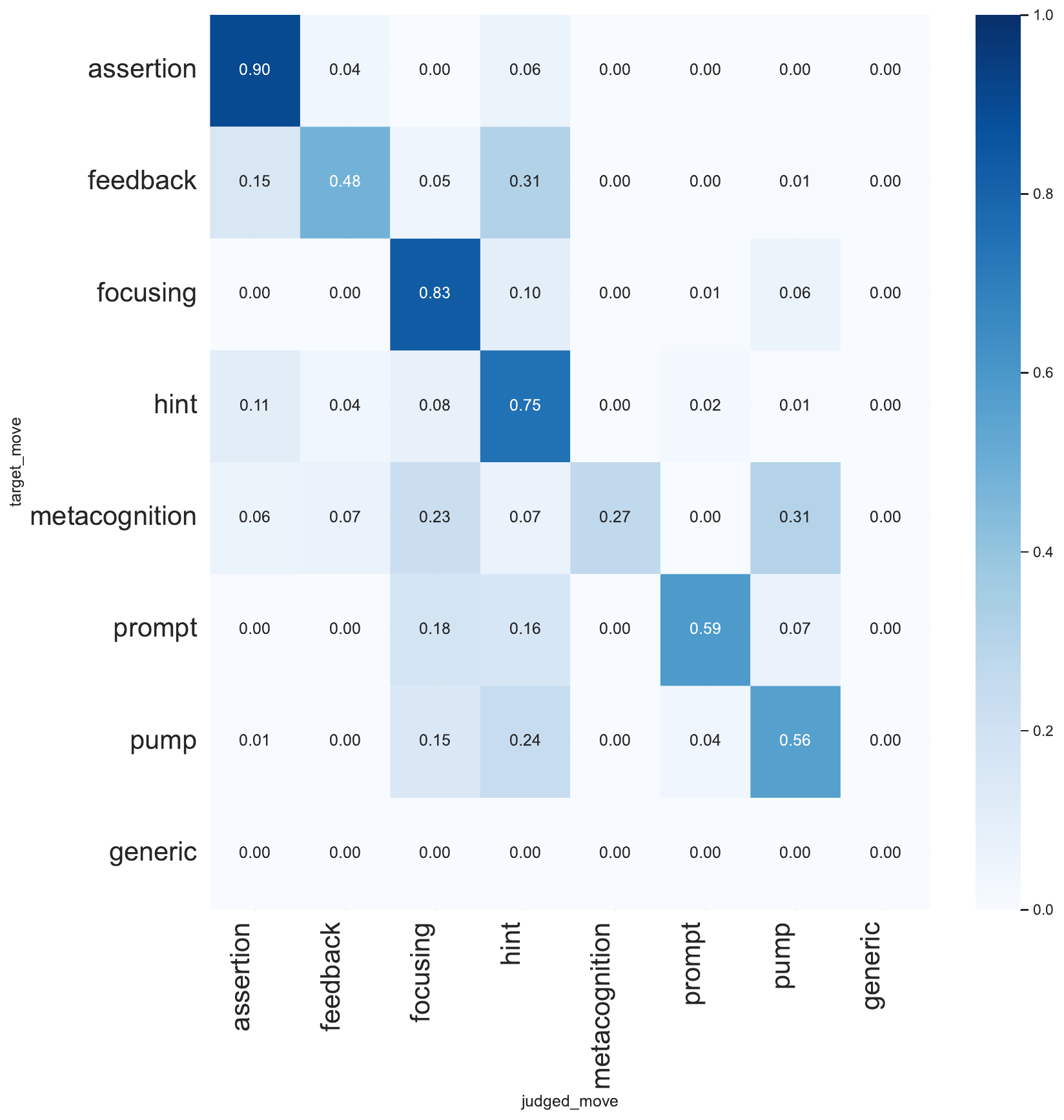}
        \caption{$\alpha=1.0$}
        \label{fig:confusion-alpha-1}
    \end{subfigure}
    \hfill
    \begin{subfigure}[t]{0.48\textwidth}
        \centering
        \includegraphics[width=\linewidth]{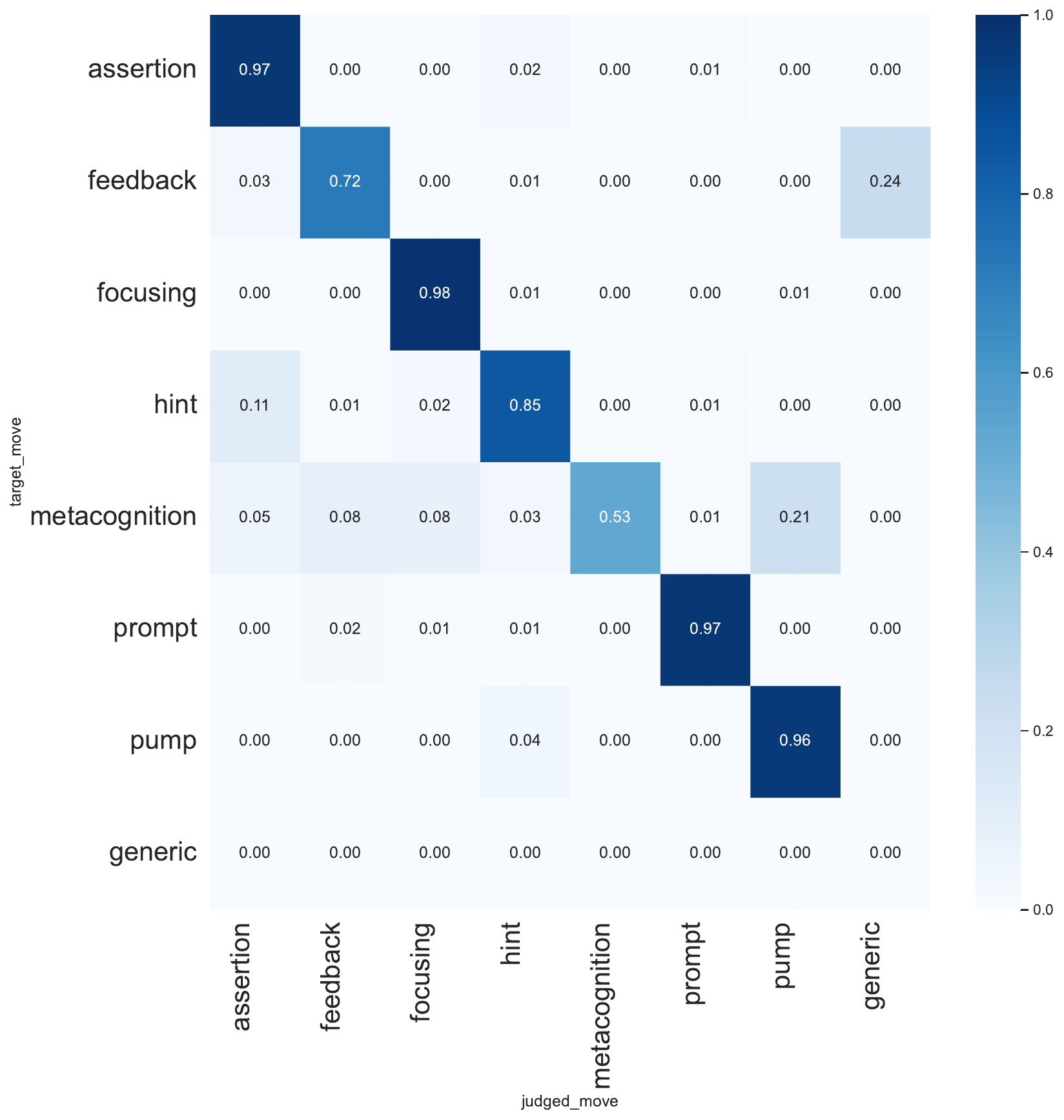}
        \caption{$\alpha=1.5$}
        \label{fig:confusion-alpha-15}
    \end{subfigure}
    \caption{
    Tutor-move confusion matrices on $D_{\mathrm{ablate}}$ at two steering strengths. Rows correspond to target moves and columns correspond to judge-predicted moves. Increasing $\alpha$ from 1.0 to 1.5 reduces many off-diagonal confusions and improves target-move specificity.
    }
    \label{fig:confusion-alpha}
\end{figure*}

Figure~\ref{fig:confusion-alpha} shows tutor-move confusion matrices on $D_{\mathrm{abl}}$ at two steering strengths. At the lower setting, $\alpha=1.0$, several target moves are still confused with neighbouring pedagogical behaviours. For example, Feedback reaches only a 48\% target-hit rate and is frequently judged as Assertion or Hint/Funnelling, while Prompt and Pump are also often confused with Focusing or Hint/Funnelling. Increasing the steering strength to $\alpha=1.5$ substantially sharpens the diagonal of the confusion matrix: Assertion, Focusing, Prompt, and Pump all exceed 95\% target-hit rate, and Feedback improves from 48\% to 72\%. However, some ambiguity remains, especially for Feedback and Metacognition, motivating the subtractive composition experiments in Section~\ref{appen:move-combos}. In particular, the subtraction setting tests whether removing confusable move directions can further improve moves specificity when a target direction overlaps with nearby tutor moves.

\section{Move Combinations}
\label{appen:move-combos}

\subsection{Subtraction}
At $\alpha=1$, feedback steering achieved a 48\% target-hit rate. 
At this $\alpha$ level, the steering was not strong enough and the feedback was often confused with nearby tutor moves such as assertion, hinting, pumping, and focusing. After subtracting the main confusable move directions, the feedback hit rate increased to 73\%, suggesting that subtraction improves the specificity of the feedback vector while preserving its intended behaviour. Table~\ref{tab:feedback-confusion-examples} shows representative examples before and after subtraction.
\begin{table*}[t]
\centering
\small
\setlength{\tabcolsep}{5pt}
\renewcommand{\arraystretch}{1.18}
\begin{tabular}{p{0.38\linewidth} p{0.16\linewidth} p{0.32\linewidth} p{0.12\linewidth}}
\toprule
\multicolumn{2}{c}{\textbf{Feedback}} &
\multicolumn{2}{c}{\textbf{Feedback $-$ Assertion $-$ Pump $-$ Hint $-$ Focusing}} \\
% \cmidrule(lr){1-2} \cmidrule(lr){3-4}
\textbf{Tutor Response} & \textbf{Judge} & \textbf{Tutor Response} & \textbf{Judge} \\
\midrule

That is correct. The viscosity depends on the temperature and gas content. &
Assertion &
That's right. &
Feedback \\

\midrule

That's a great start. You correctly identified the greenhouse effect. Now, consider how the thick cloud cover interacts with the heat. &
Hint/Funneling &
That is not quite right. &
Feedback \\

\midrule

What else could be going on here? &
Pump &
That's not quite right. You are describing variation, but you haven't identified the specific reason yet. &
Feedback \\

\midrule

How did you decide where they fit? &
Focusing &
That's right. &
Feedback \\

\bottomrule
\end{tabular}
\caption{
Examples of feedback-targeted generations before and after subtracting confusable move directions. The examples include both positive and negative feedback corrections.
}
\label{tab:feedback-confusion-examples}
\end{table*}

\subsection{Addition}
We tested an additive setting in which two tutor-move vectors are applied together at inference time. Because our tutor-move judge assigns exactly one label to each response, a successful combination is reflected when the response is judged as one of the two moves being combined. Table~\ref{tab:additive-move-examples} shows qualitative examples where the generated tutor response reflects one of the intended added move directions.

\begin{table*}[t]
\centering
\footnotesize
\setlength{\tabcolsep}{3pt}
\renewcommand{\arraystretch}{1.18}
\begin{tabular}{p{0.18\linewidth} p{0.67\linewidth} p{0.12\linewidth}}
\toprule
\textbf{Tutor Moves Added} & \textbf{Tutor Response} & \textbf{Judge Label} \\
\midrule

\multirow{2}{*}{Assertion + Feedback} &
That is correct: shield volcanoes are the least explosive, stratovolcanoes are the most explosive, and cinder cones and lava domes fall in between. &
Assertion \\
\cmidrule(lr){2-3}
&
You calculated average speed, but the question asks for the average time, which requires determining Diego's full race time of 5 minutes. &
Hint/Funneling \\

\midrule

\multirow{2}{*}{Feedback + Focusing} &
That is a good summary of the extremes; let's check where cinder cones and lava domes fit by considering their lava viscosity. &
Hint/Funneling \\
\cmidrule(lr){2-3}
&
You correctly identified that the question asks for average speed, but your calculation of the total time assumes Diego finished the race in the same time he ran half. Since Diego was tripped, how did you determine the total time for both racers? &
Focusing \\

\midrule

\multirow{2}{*}{Feedback + Pump} &
That is a good hypothesis to test; how do the eruption styles of cinder cones and lava domes compare to those two extremes? &
Pump \\
\cmidrule(lr){2-3}
&
You correctly identified that the question asks for average speed, but you misinterpreted the condition regarding Diego finishing the race. Remember, the problem states he ``can't finish,'' so you need to calculate his time as if he had completed the full distance at that same speed. &
Hint/Funneling \\

\midrule

\multirow{2}{*}{Focusing + Hint} &
You placed cinder cones and lava domes in the middle of explosiveness; what specific property of their lava or gas content leads you to that ranking? &
Focusing \\
\cmidrule(lr){2-3}
&
You correctly identified that the problem asks for average speed, but did you account for Diego's full race time when calculating the total time used in your division? &
Hint/Funneling \\

\midrule

\multirow{2}{*}{Hint + Assertion} &
Cinder cones typically have moderate explosivity due to gas-rich ash, while lava domes are highly explosive because their extremely thick lava traps gases near the vent. &
Assertion \\
\cmidrule(lr){2-3}
&
You need to calculate Diego's \emph{projected} time to finish the full block based on his speed, not just the time he actually ran. &
Hint/Funneling \\

\bottomrule
\end{tabular}
\caption{
Qualitative examples from additive tutor-move steering. Since the judge assigns one label per response, combined move vectors are reflected when the generated response is classified as one of the two intended moves.
}
\label{tab:additive-move-examples}
\end{table*}
% \section{BiPO vs PIVOT}
\section{Comparison to Offline Preference Optimisation}
\label{appen:bipo-comparison}

We compare PIVOT with an offline preference-optimisation baseline adapted from BiPO \cite{cao2024personalized}. This offline baseline learns steering vectors from a fixed set of preference pairs, whereas PIVOT constructs preferences online from the model's own target and confusable non-target generations. For a fair comparison, we trained BiPO-style vectors using the preference pairs generated in all rounds of PIVOT's online training, learned vectors for all tutor moves across layers 6-28, and used the same number of optimisation steps as PIVOT.

\begin{figure}[t]
    \centering
    \includegraphics[width=\columnwidth]{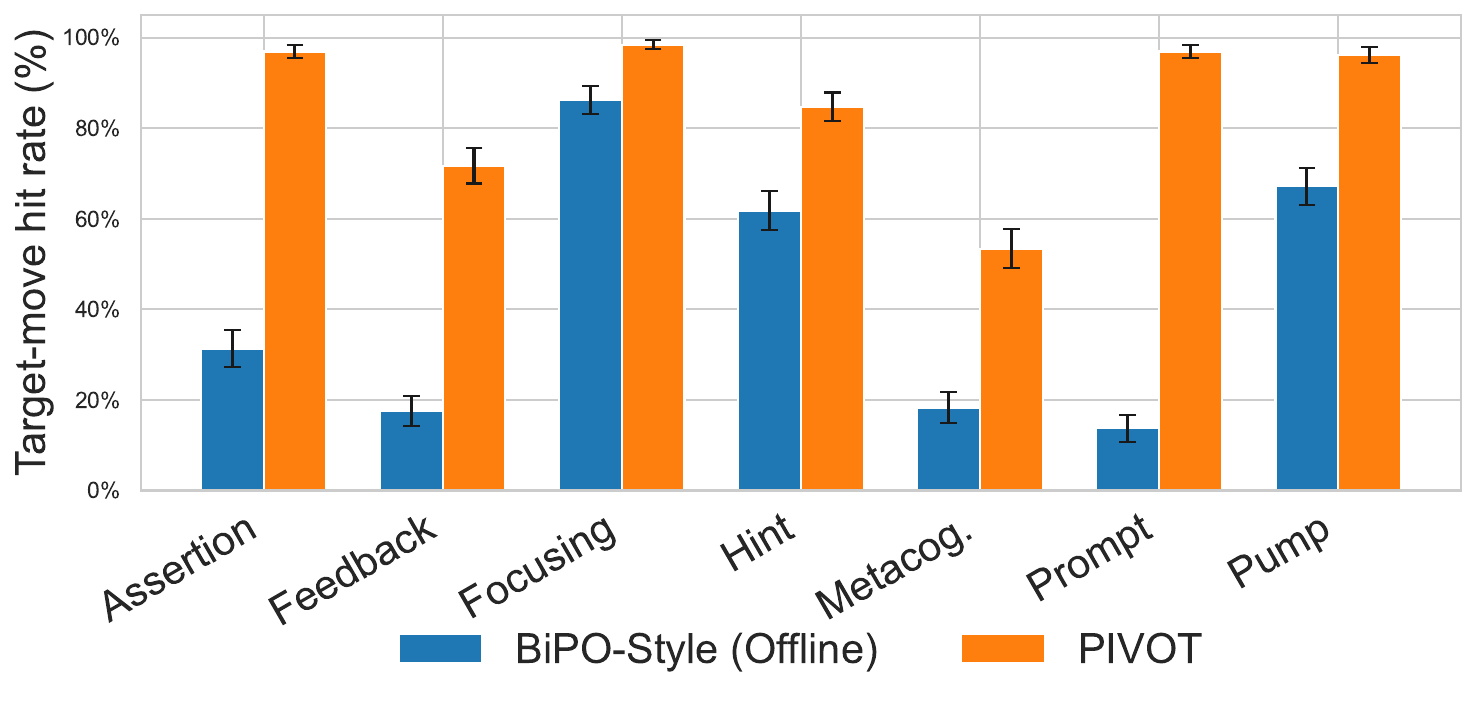}
    \caption{Target-move hit rates for the offline BiPO baseline and PIVOT on $\mathcal{D}_{\mathrm{ablate}}$.}
    \label{fig:bipo-vs-pivot}
\end{figure}

Figure~\ref{fig:bipo-vs-pivot} shows that PIVOT achieves substantially higher target-move hit rates than the offline BiPO-style baseline across all tutor moves at $\alpha=1.5$ . The gains are especially large for Assertion, Feedback, Prompt, and Metacognition, while Focusing is strong for both methods but still improves under PIVOT. These results suggest that online preference construction is important for tutor-move steering: by repeatedly sampling from the current steered model and upweighting its confusions, PIVOT learns directions that better separate target moves from nearby pedagogical behaviours than vectors trained from a fixed offline preference set.

\section{Qwen-4B Evaluation}
\label{appen:Qwen4b}
To test whether PIVOT extends to smaller models, we trained Qwen3.5-4B using the same training parameters as the 9B model and evaluated steering at $\alpha=1.5$. As shown in Table~\ref{tab:qwen4b-results}, the smaller model achieves strong target-move control for most moves, with especially high hit rates for Focusing (95.4\%), Assertion (87.0\%), Hint (85.5\%), Pump (84.7\%), and Feedback (80.2\%).

Relevance remains high for Hint and Prompt, but drops for Metacognition and Pump, suggesting that some reflective or open-ended moves are harder to steer reliably in the smaller model. Additional $\alpha$ tuning may be required to improve Metacognition and Pump performance. Fluency is generally strong for Assertion, Feedback, and Hint, although Focusing, Pump, and Metacognition show lower fluency. This indicates a trade-off between move control and response quality for some categories in a smaller model such as Qwen3.5-4B.

Overall, these results suggest that PIVOT remains effective on a smaller 4B model, while also highlighting which tutor moves are more sensitive to model capacity. Future work could further characterise the behaviour of trained steering vectors in smaller models.

\begin{table}[t]
\centering
\small
\setlength{\tabcolsep}{5pt}
\renewcommand{\arraystretch}{1.12}
\resizebox{\columnwidth}{!}{%
\begin{tabular}{lcccc}
\toprule
\textbf{Move} & \textbf{N} & \textbf{Target Hit (\%)} & \textbf{Relevance (\%)} & \textbf{Fluency (\%)} \\
\midrule
Assertion     & 132 & $87.0 \pm 2.9$ & $80.2 \pm 3.5$ & $91.6 \pm 2.4$ \\
Feedback      & 132 & $80.2 \pm 3.5$ & $72.5 \pm 3.9$ & $91.6 \pm 2.4$ \\
Focusing      & 132 & $95.4 \pm 1.8$ & $80.2 \pm 3.5$ & $60.3 \pm 4.3$ \\
Hint          & 132 & $85.5 \pm 3.1$ & $90.1 \pm 2.6$ & $90.8 \pm 2.5$ \\
Metacognition & 132 & $61.8 \pm 4.3$ & $58.8 \pm 4.3$ & $71.0 \pm 4.0$ \\
Prompt        & 132 & $75.6 \pm 3.8$ & $87.8 \pm 2.9$ & $75.6 \pm 3.8$ \\
Pump          & 132 & $84.7 \pm 3.2$ & $64.9 \pm 4.2$ & $71.8 \pm 3.9$ \\
\bottomrule
\end{tabular}%
}
\caption{
Qwen3.5-4B steering results at $\alpha=1.5$ using the same training setup as the 9B model. Values report percentages with confidence intervals.
}
\label{tab:qwen4b-results}
\end{table}

\section{Ablation Test-set Evaluation}
\label{appen:abl-testset}

\begin{figure}[t]
    \centering
    \includegraphics[width=0.9\columnwidth]{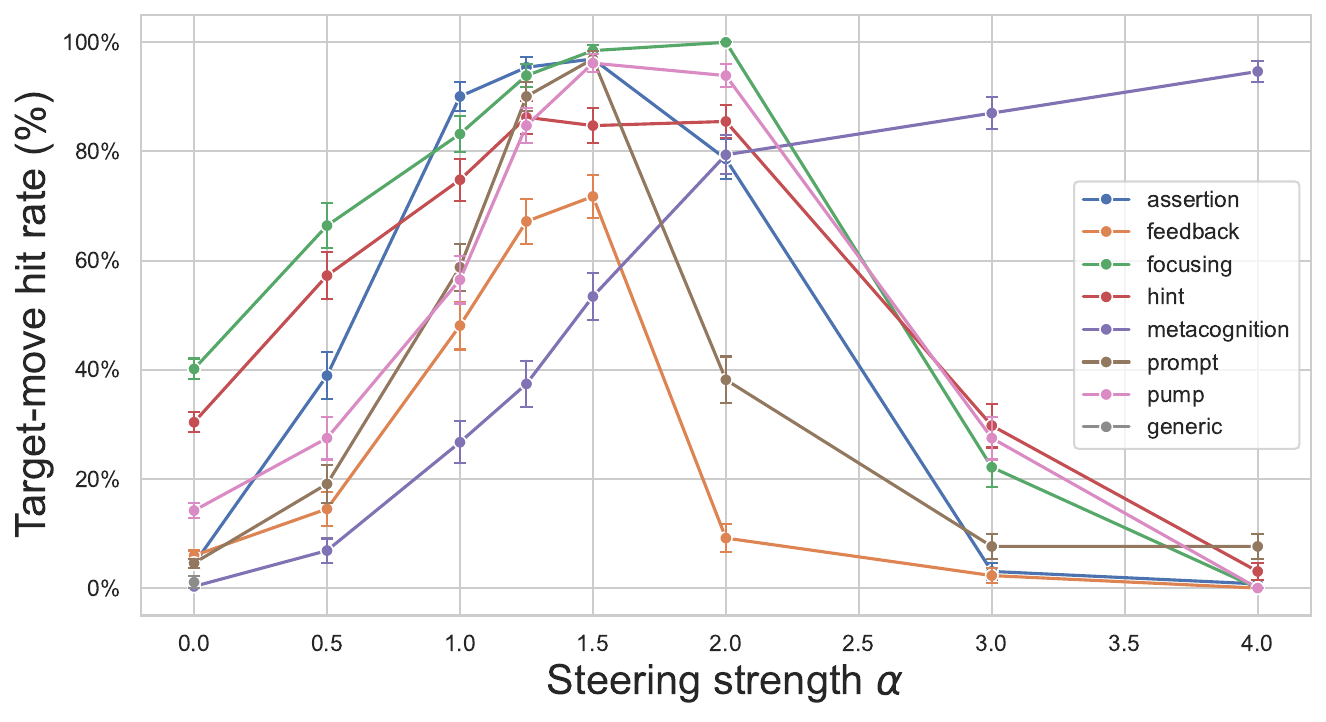}
    \caption{Target-move hit rate across steering strengths on $\mathcal{D}_{\mathrm{ablate}}$.}
    \label{fig:dablate-alpha-sweep}
\end{figure}

\begin{figure}[t]
    \centering
    \includegraphics[width=0.9\columnwidth]{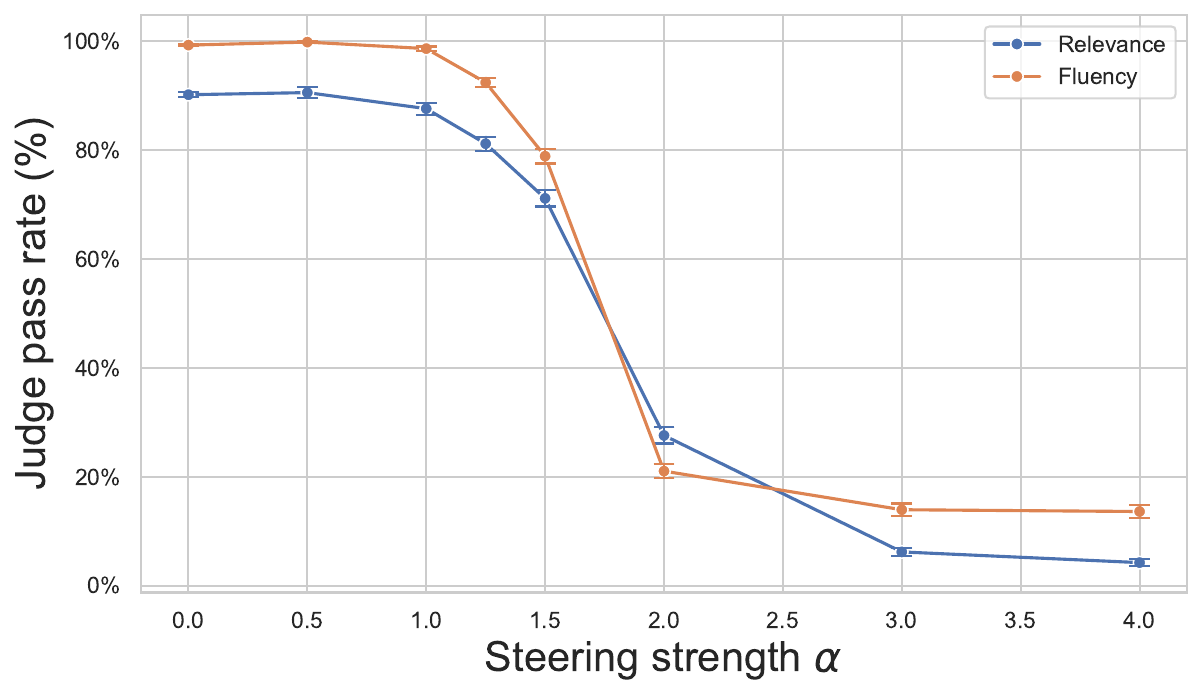}
    \caption{Relevance and fluency pass rates across steering strengths on $\mathcal{D}_{\mathrm{ablate}}$.}
    \label{fig:dablate-quality-sweep}
\end{figure}

We repeated the steering-strength sweep on $\mathcal{D}_{\mathrm{ablate}}$ to verify that this smaller set follows the same controllability--quality pattern as $\mathcal{D}_{\mathrm{sweep}}$. As shown in Figure~\ref{fig:dablate-alpha-sweep}, target-move hit rates increase from low steering strengths, peak around moderate values, and then degrade at high $\alpha$. Figure~\ref{fig:dablate-quality-sweep} shows the same quality trade-off observed on $\mathcal{D}_{\mathrm{sweep}}$: relevance and fluency remain high at small to moderate strengths but drop sharply once steering becomes too strong. These similar trends justify using $\mathcal{D}_{\mathrm{ablate}}$ for compute-intensive ablations.

\section{More Implementation Details}

Table~\ref{tab:implementation-details} summarizes the main optimization hyperparameters used to learn the steering vectors. We use a norm penalty and a hard maximum vector norm to prevent the learned vectors from becoming overly large or producing overly aggressive interventions.

\begin{table}[t]
\centering
\small
\setlength{\tabcolsep}{6pt}
\renewcommand{\arraystretch}{1.12}
\begin{tabular}{lc}
\toprule
\textbf{Hyperparameter} & \textbf{Value} \\
\midrule
Learning rate & $4 \times 10^{-4}$ \\
Training steps per round & 50 \\
Rounds & 5 \\
DPO loss $\beta$ & 0.1 \\
Norm penalty & 0.008 \\
Max vector norm & 6 \\
\bottomrule
\end{tabular}
\caption{Hyperparameters used for preference-based steering-vector learning.}
\label{tab:implementation-details}
\end{table}
\section{User Study}

\label{appen:user-study}
\subsection{Student Agent}
We used a simulated student agent to generate short student replies during the tutoring interaction. The prompt instructs the model to remain in the student role, respond briefly, and avoid becoming too capable too quickly.

\newtcblisting{pinkpromptbox}[1]{
    title=\textbf{#1},
    colback=pink!6,
    colframe=pink!35,
    coltitle=black,
    fonttitle=\bfseries,
    boxrule=0.5pt,
    arc=1mm,
    left=6pt,
    right=6pt,
    top=5pt,
    bottom=5pt,
    breakable,
    listing only,
    listing options={
        basicstyle=\ttfamily\footnotesize,
        breaklines=true,
        breakatwhitespace=true,
        columns=fullflexible,
        keepspaces=true,
        showstringspaces=false
    }
}

\begin{pinkpromptbox}{Student Agent Prompt}
You are simulating a math student in a tutoring study.
Respond only as the student, in one or two sentences.
Do not become too capable too quickly.

Scenario: {scenario_name}
Problem: {problem}
Initial student answer: {initial}

Conversation so far:
{history}

Student reply:
\end{pinkpromptbox}

\subsection{Participants Background}
\begin{figure}[t]
    \centering

    \begin{subfigure}[t]{0.31\columnwidth}
        \centering
        \includegraphics[width=\linewidth]{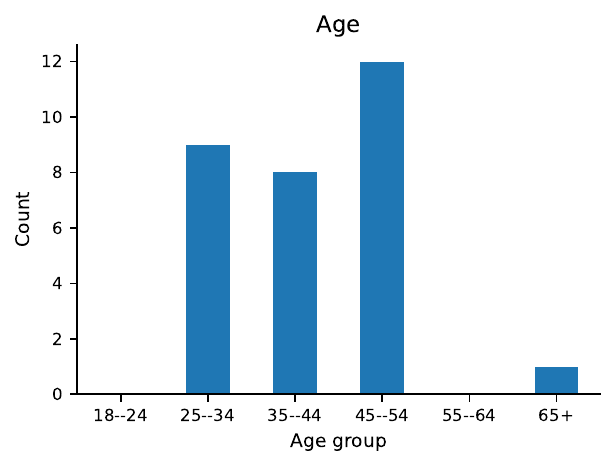}
        \caption{Age}
    \end{subfigure}
    \hfill
    \begin{subfigure}[t]{0.31\columnwidth}
        \centering
        \includegraphics[width=\linewidth]{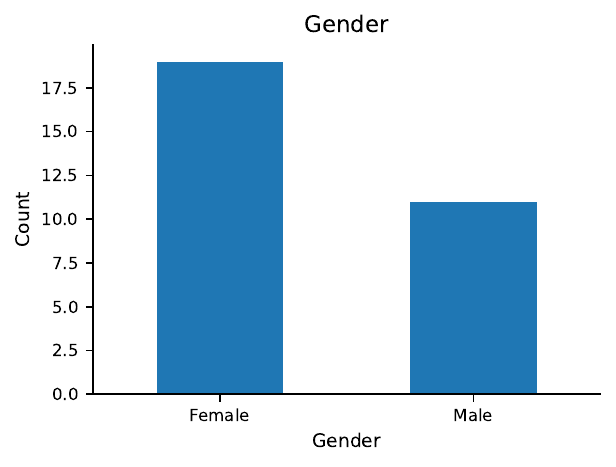}
        \caption{Gender}
    \end{subfigure}
    \hfill
    \begin{subfigure}[t]{0.31\columnwidth}
        \centering
        \includegraphics[width=\linewidth]{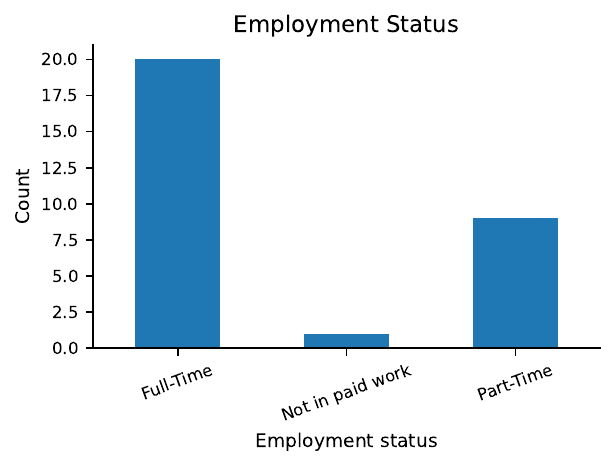}
        \caption{Employment}
    \end{subfigure}

    \vspace{0.2cm}

    \begin{subfigure}[t]{0.48\columnwidth}
        \centering
        \includegraphics[width=\linewidth]{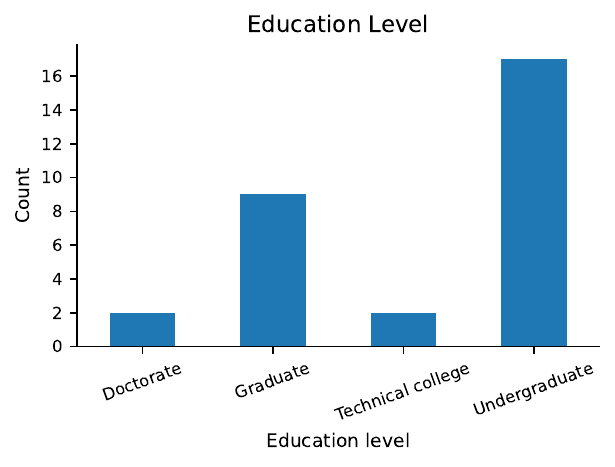}
        \caption{Education level}
    \end{subfigure}
    \hfill
    \begin{subfigure}[t]{0.48\columnwidth}
        \centering
        \includegraphics[width=\linewidth]{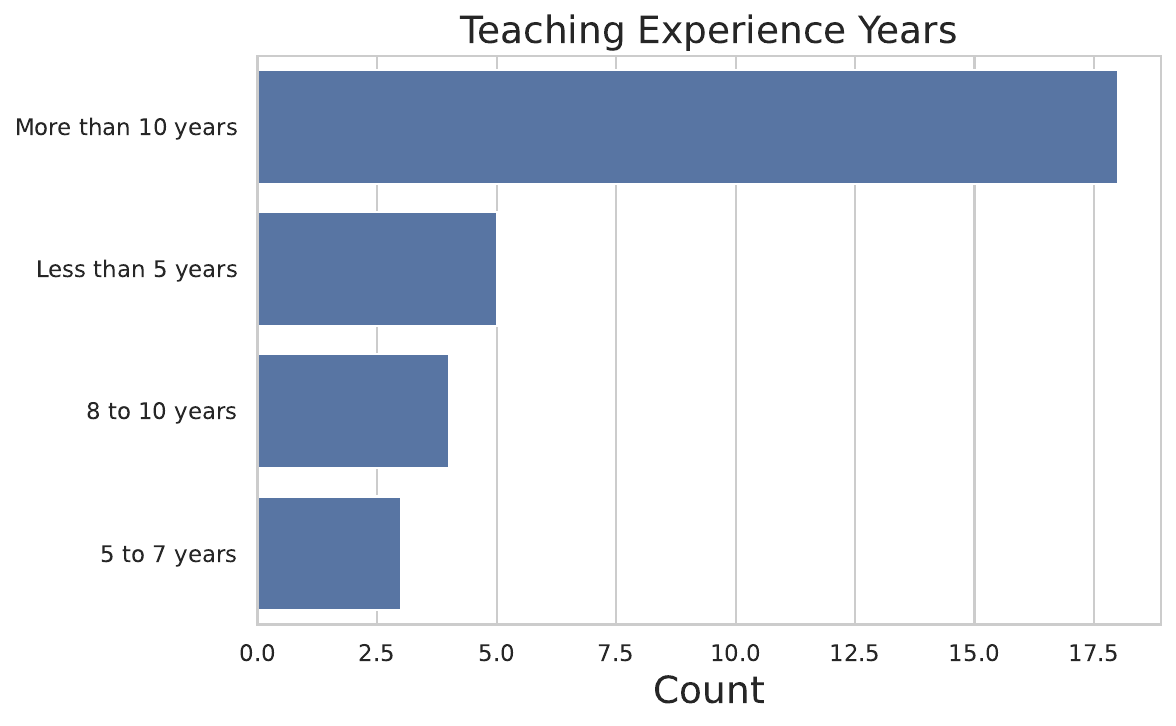}
        \caption{Teaching experience}
    \end{subfigure}

    \vspace{0.2cm}

    \begin{subfigure}[t]{0.48\columnwidth}
        \centering
        \includegraphics[width=\linewidth]{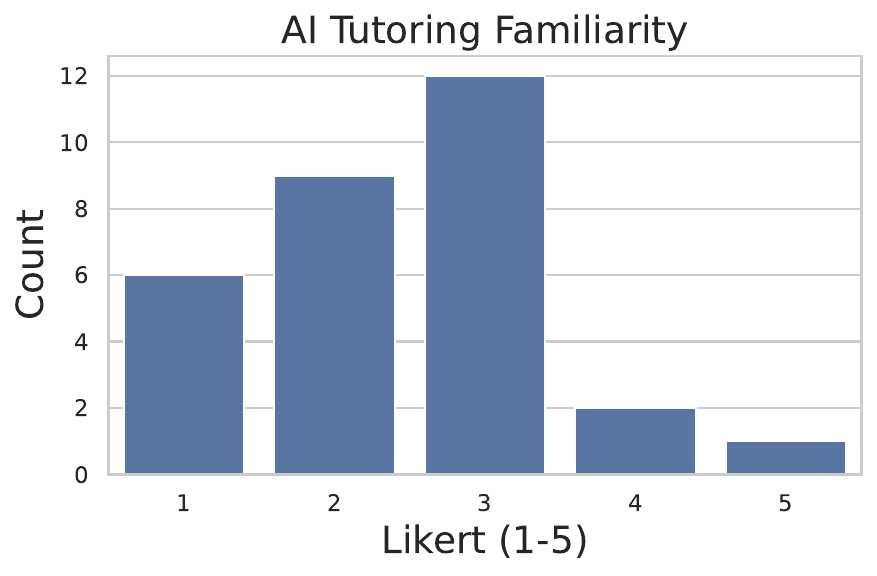}
        \caption{AI tutoring familiarity}
    \end{subfigure}
    \hfill
    \begin{subfigure}[t]{0.48\columnwidth}
        \centering
        \includegraphics[width=\linewidth]{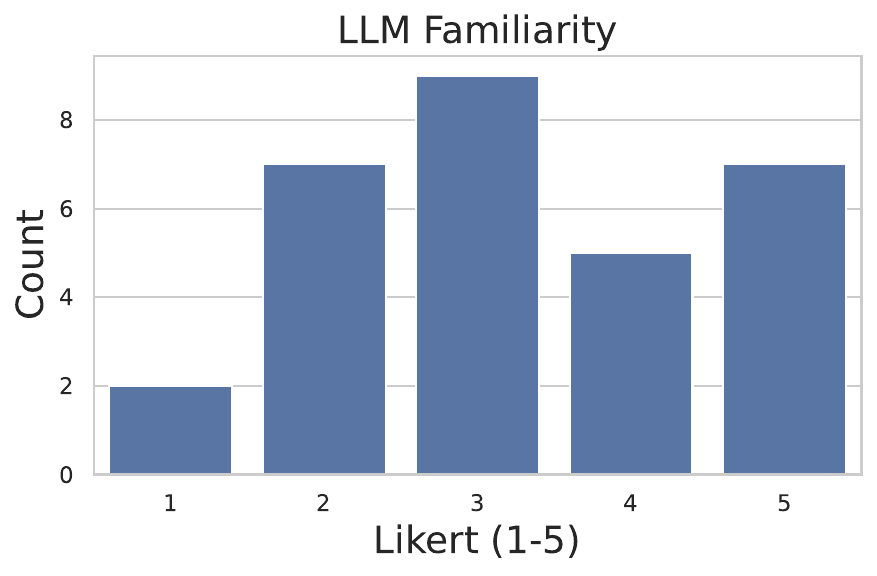}
        \caption{LLM familiarity}
    \end{subfigure}

    \vspace{0.2cm}

    \begin{subfigure}[t]{0.48\columnwidth}
        \centering
        \includegraphics[width=\linewidth]{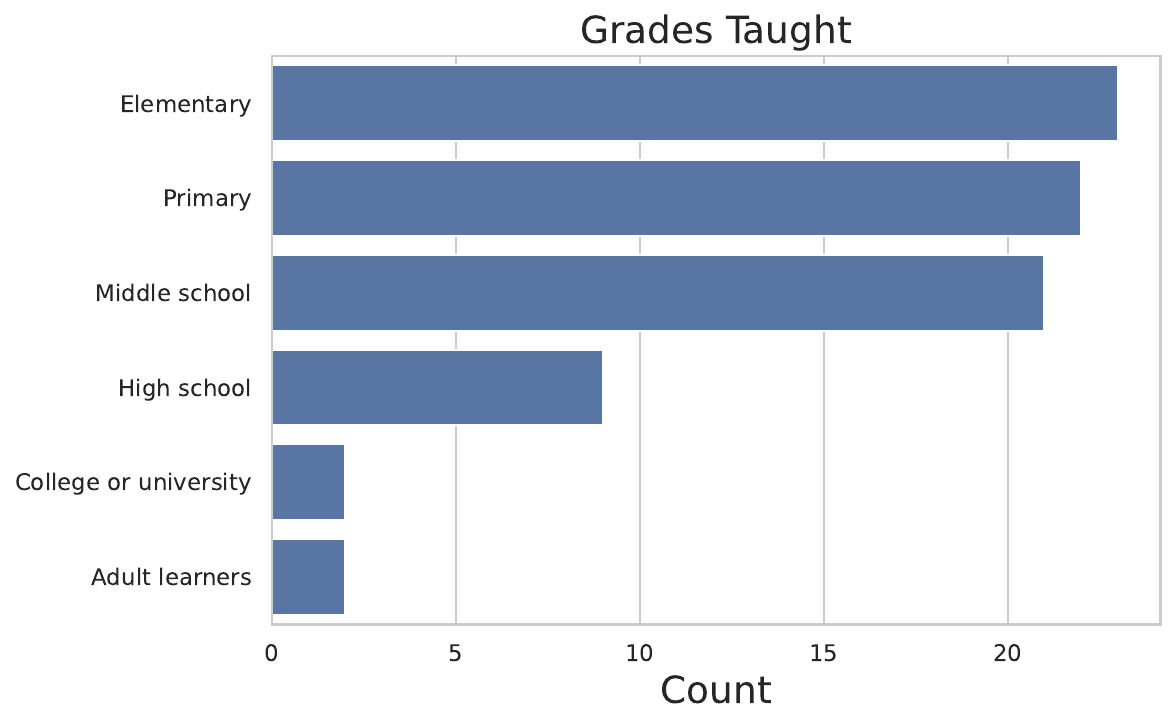}
        \caption{Grades taught}
    \end{subfigure}
    \hfill
    \begin{subfigure}[t]{0.48\columnwidth}
        \centering
        \includegraphics[width=\linewidth]{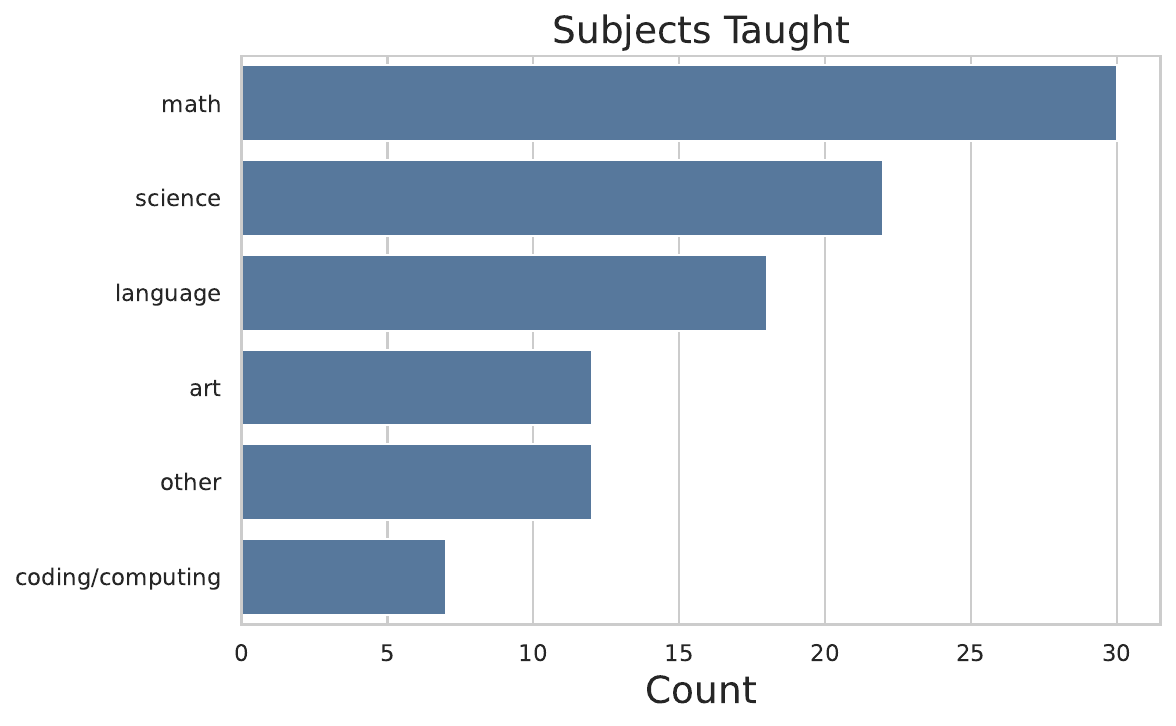}
        \caption{Subjects taught}
    \end{subfigure}

    \caption{Demographics of study participants.}
    \label{fig:demographics}
\end{figure}

Figure~\ref{fig:demographics} summarises the demographics of the 30 teacher participants. Participants covered a range of age groups, with most between 25 and 54 years old, and included 19 female and 11 male participants. Most were employed full-time and held an undergraduate or graduate degree. All participants reported mathematics teaching experience, with many also teaching science, language, art, or computing. Participants primarily taught elementary, primary, and middle-school students, and most had more than 10 years of teaching experience. Familiarity with AI tutoring tools was mostly moderate, while LLM familiarity was more varied. The median completion time was 18 minutes, corresponding to an average hourly rate of \pounds14.86.

% \begin{figure}[t]
%     \centering

%     \begin{subfigure}[t]{0.5\columnwidth}
%         \centering
%         \includegraphics[width=\linewidth]{figures/pre-survey/ai_tutoring_familiarity_count.pdf}
%         \caption{AI Tutoring Familiarity}
%     \end{subfigure}
%     \hfill
%     \begin{subfigure}[t]{0.5\columnwidth}
%         \centering
%         \includegraphics[width=\linewidth]{figures/pre-survey/grades_taught_count.pdf}
%         \caption{Grades Taught}
%     \end{subfigure}
%     \hfill
%     \begin{subfigure}[t]{0.5\columnwidth}
%         \centering
%         \includegraphics[width=\linewidth]{figures/pre-survey/llm_familiarity_count.pdf}
%         \caption{LLMs Familiarity}
%     \end{subfigure}

%     \vspace{0.15cm}

%     \begin{subfigure}[t]{0.5\columnwidth}
%         \centering
%         \includegraphics[width=\linewidth]{figures/pre-survey/subjects_taught_count.pdf}
%         \caption{Subjects taught}
%     \end{subfigure}
%     \hfill
%     \begin{subfigure}[t]{0.5\columnwidth}
%         \centering
%         \includegraphics[width=\linewidth]{figures/pre-survey/teaching_experience_years_count.pdf}
%         \caption{Education level}
%     \end{subfigure}

%     \caption{Demographics of study participants.}
%     \label{fig:demographics}
% \end{figure}

\newtcolorbox{bluestudybox}[1]{
    title=\textbf{#1},
    colback=cyan!5,
    colframe=cyan!35,
    coltitle=black,
    fonttitle=\bfseries,
    boxrule=0.5pt,
    arc=1mm,
    left=6pt,
    right=6pt,
    top=5pt,
    bottom=5pt,
    breakable
}

\subsection{Study Context and Questionnaires}

\begin{bluestudybox}{Explanatory Statement and Consent}
Dear participant,

Thank you for participating in our study on controllable AI tutoring. We are very grateful for your participation and your invaluable insight. Please read this Explanatory Statement in full before proceeding. If you would like further information regarding any aspect of this project, please contact us using the email address provided below.

We are a group of researchers dedicated to improving education through technology. The goal of this study is to evaluate whether language model tutors can be made more controllable for teachers and tutors. In particular, we are interested in whether teachers can use simple controls to guide an AI tutor toward different tutoring behaviours, such as giving broad guidance, narrowing the student's attention, giving a hint, asking a reflective question, or providing direct explanation.

\begin{itemize}
    \item All personal information will be kept confidential and anonymised. Only demographic information is recorded, and it will be reported only in aggregate form to prevent identifying any individual participant.
    \item You may withdraw at any time. Any data you have provided up to that point will be destroyed.
    \item All data will be collected, stored securely, and reported in accordance with applicable data protection law.
    \item Only anonymised or aggregated data may be used in future research, subject to ethics approval, and made available to other researchers for further analysis and verification.
    \item Only the principal investigator and the designated researchers will have access to the original data under strict confidentiality. Results from the project may be published in conference papers and/or journal articles, but no personal data will be shared.
    \item Personal data will be stored for 5 years from the date of collection. During this period, participants have the right to access their data and inquire about its processing. To exercise this right, contact the Principal Investigator.
\end{itemize}

By participating in this survey, you agree that your data may be used for scientific purposes.
\end{bluestudybox}

\begin{bluestudybox}{What You Will Do in This Study}
In this study, you will:

\begin{itemize}
    \item You must have experience teaching elementary/middle school math.
    \item Learn about simple tutor response types.
    \item Complete a short practice activity.
    \item Try move controls that change how the AI tutor responds.
    \item Use the controls on a student answer.
    \item Compare your controlled tutor conversation with a neutral AI-tutor conversation.
    \item Answer a short reflection about whether the controls were useful.
\end{itemize}

The goal is to understand whether teachers and tutors can use these controls to guide an AI tutor toward the response they want.

The study should take approximately \textbf{20 to 30 minutes}. Please complete it carefully and in full. Incomplete or unserious submissions may be discarded.
\end{bluestudybox}

\begin{bluestudybox}{Study Context}
Large language models are increasingly used to support education through tutoring, feedback, and question generation. They can often produce fluent and useful responses, but they are not always easy to control.

The goal of tutoring is not only to give a correct answer, but also to respond in the right way for the situation. For example, a tutor may:

\begin{itemize}
    \item Ask the student to explain more.
    \item Give a small hint.
    \item Focus attention on a mistake.
    \item Encourage reflection.
    \item Directly explain an idea.
\end{itemize}

Normally, AI tutors are guided using written instructions. However, once these instructions are given, the tutor's behaviour is difficult to control. Even with the same instructions, the AI may sometimes:

\begin{itemize}
    \item Give too much help.
    \item Be too vague.
    \item Ask the wrong kind of question.
    \item Respond in a way that is not helpful for learning.
\end{itemize}

In this study, we compare two approaches using the same tutoring instructions:

\begin{itemize}
    \item A standard AI tutor that only follows the written prompt.
    \item \textbf{A controllable AI tutor} where tutoring behaviours can be \textbf{adjusted manually} using simple tutor-move controls.
\end{itemize}

The goal is to understand whether manual control can help teachers guide AI tutor behaviour more effectively.
\end{bluestudybox}

\begin{bluestudybox}{Pre-task Questionnaire}
\begin{enumerate}[leftmargin=*]
    \item \textbf{Do you have experience with elementary/middle school math?}
    \begin{itemize}[leftmargin=*]
        \item Yes
        \item No
    \end{itemize}

    \item \textbf{Years of teaching or tutoring experience}
    \begin{itemize}[leftmargin=*]
        \item Less than 5 years
        \item 5 to 7 years
        \item 8 to 10 years
        \item More than 10 years
    \end{itemize}

    \item \textbf{Grades or levels taught}
    \begin{itemize}[leftmargin=*]
        \item Primary
        \item Elementary
        \item Middle school
        \item High school
        \item College or university
        \item Adult learners
        \item Other
    \end{itemize}

    \item \textbf{Subjects taught}
    \begin{itemize}[leftmargin=*]
        \item Math
        \item Language
        \item Science
        \item Coding/computing
        \item Art
        \item Other
    \end{itemize}

    \item \textbf{Familiarity with AI tutoring tools}
    \begin{itemize}[leftmargin=*]
        \item 1 = Not at all familiar
        \item 2 = Slightly familiar
        \item 3 = Moderately familiar
        \item 4 = Very familiar
        \item 5 = Extremely familiar
    \end{itemize}

    \item \textbf{Familiarity with Large Language Models (LLMs)}
    \begin{itemize}[leftmargin=*]
        \item 1 = Not at all familiar
        \item 2 = Slightly familiar
        \item 3 = Moderately familiar
        \item 4 = Very familiar
        \item 5 = Extremely familiar
    \end{itemize}
\end{enumerate}
\end{bluestudybox}

\begin{bluestudybox}{Post-task Questionnaire}
\textbf{Step 1: Likert Questionnaire}

Prompt: How much do you agree with the following, or to what extent do you agree?

Scale:
\begin{itemize}
    \item 1 -- Strongly disagree
    \item 2 -- Disagree
    \item 3 -- Neither agree nor disagree
    \item 4 -- Agree
    \item 5 -- Strongly agree
\end{itemize}

Items:
\begin{itemize}
    \item I felt able to control the tutor's overall response strategy.
    \item I found the effect of the move controls clear.
    \item It was easy to set the controls for the tutoring behaviour I wanted.
    \item I felt the move controls helped me provide the kind of guidance and help I intended for this student.
    \item I felt the move controls helped me guide the student without giving away the answer too early.
    \item I found the controlled responses coherent with the student's previous messages.
    \item I found the controlled responses respectful and supportive, without being dismissive, rude, overly harsh, or unnatural.
    \item I found this method useful for guiding how the AI tutor helped the student.
    \item I would use this kind of controllability method in future tutoring practice.
\end{itemize}

These Likert items are shown in randomised order per participant, and all items are required before continuing.

\textbf{Step 2: Overall Preference}

Prompt: Which conversation did you prefer overall?

Options:
\begin{itemize}
    \item Neutral conversation
    \item Controlled conversation
    \item No preference
\end{itemize}

Follow-up: Why do you prefer this conversation?

A selection and reason text are required before continuing.

\textbf{Step 3: Open-Ended Feedback}

Prompts:
\begin{itemize}
    \item Which move control was most useful, and why?
    \item What would make this method more useful for real tutoring?
\end{itemize}

Both responses are required before continuing.

\textbf{Step 4: Completion}

Fields shown:
\begin{itemize}
    \item Prolific redirect URL
    \item Completion code
    \item Submit button
\end{itemize}

Additional recorded field: \texttt{neutral\_vs\_steered\_difference}, from the main-task comparison response.
\end{bluestudybox}

\subsection{Move-combination Usage by Scenario}

\begin{figure}[t]
    \centering

    \begin{subfigure}[t]{0.95\columnwidth}
        \centering
        \includegraphics[width=\linewidth]{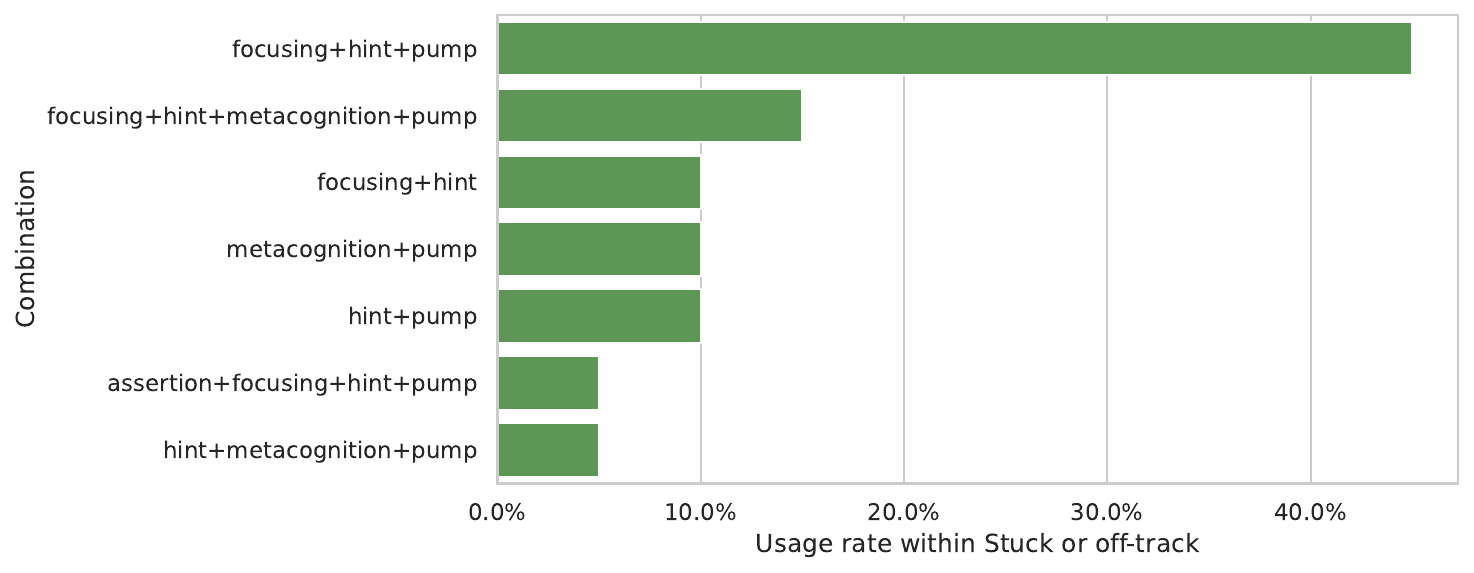}
        \caption{Stuck or off-track}
    \end{subfigure}

    \vspace{0.25cm}

    \begin{subfigure}[t]{0.95\columnwidth}
        \centering
        \includegraphics[width=\linewidth]{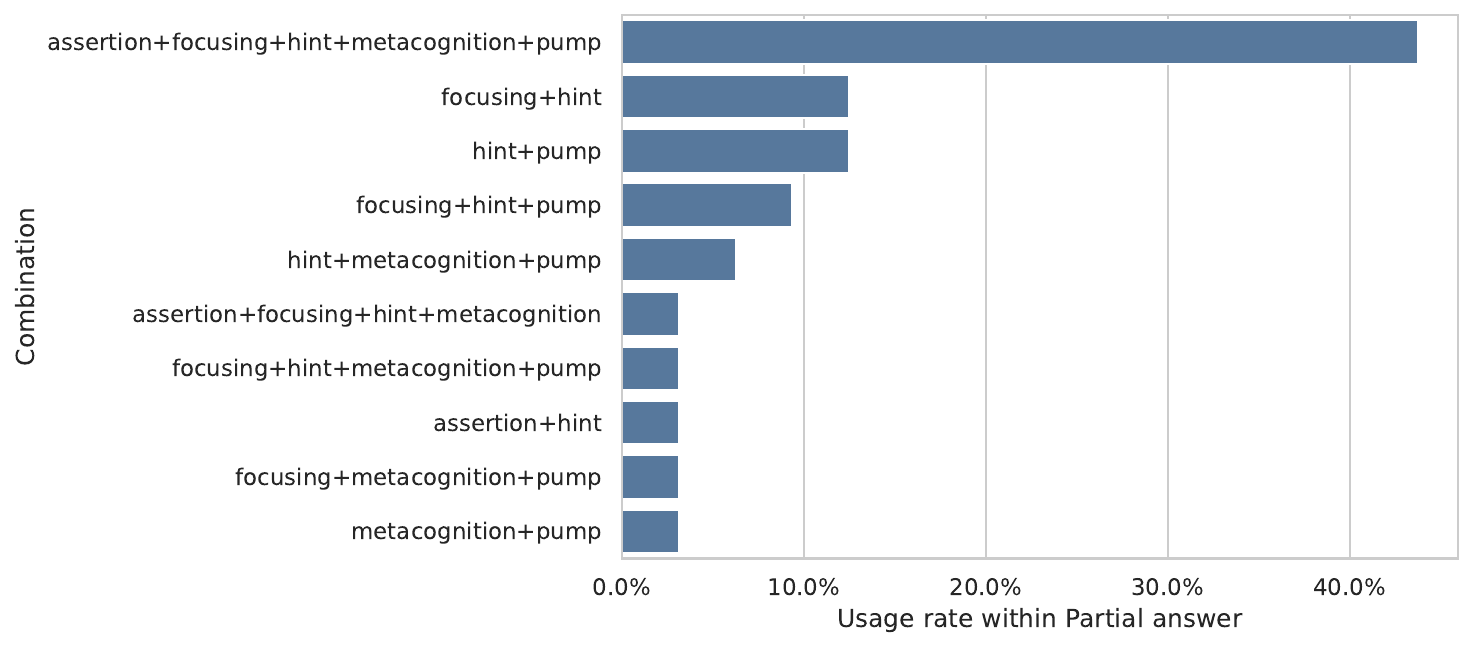}
        \caption{Partial answer}
    \end{subfigure}

    \vspace{0.25cm}

    \begin{subfigure}[t]{0.95\columnwidth}
        \centering
        \includegraphics[width=\linewidth]{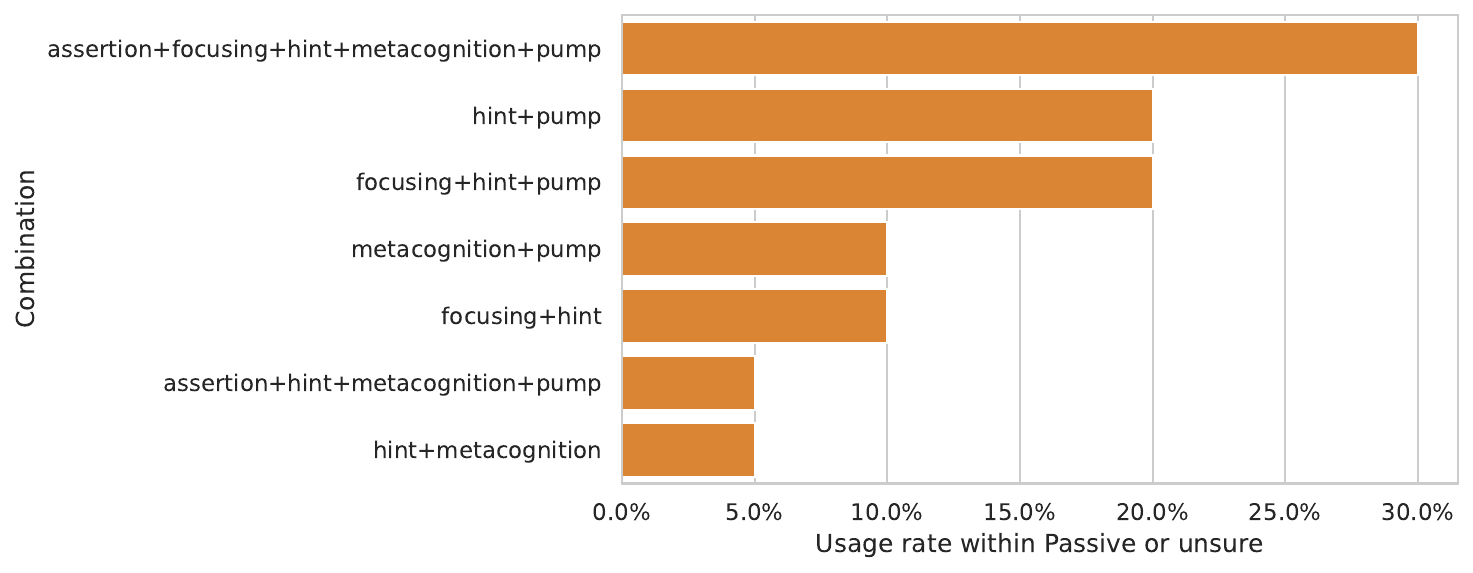}
        \caption{Passive or unsure}
    \end{subfigure}

    \caption{Usage rates of teacher-selected move-control combinations by student scenario.}
    \label{fig:move-combo-usage-scenarios}
\end{figure}

We also analysed how often teachers used different combinations of move controls in each student scenario. Figure~\ref{fig:move-combo-usage-scenarios} shows the most frequent move combinations for the \textit{stuck or off-track}, \textit{partial answer}, and \textit{passive or unsure} scenarios. Across scenarios, teachers frequently combined Hint with Focusing and Pump, suggesting that they used the controls to provide targeted guidance while still leaving room for student reasoning. For stuck or off-track students, the most common combination was Focusing+Hint+Pump, indicating a preference for narrowing attention, giving support, and eliciting further reasoning. For partial-answer and passive or unsure students, teachers more often included broader combinations with Assertion and Metacognition, suggesting that they sometimes combined direct explanation or reflection with guided support when students had already made progress or lacked confidence.

% \section{Example Appendix}
% \label{sec:appendix}
% \input{sections/appendices/data_collection}
% \input{sections/appendices/full_results}
% \input{sections/appendices/ablations}
% \input{sections/appendices/gpt-judge}
% \input{sections/appendices/data_generation}
% \input{sections/appendices/tools}
% \input{sections/appendices/user_study}
% \input{sections/appendices/data_collection}
% \input{sections/appendices/self_reflection_prompt}
% \input{sections/appendices/tool_contributions}

\end{document}